\documentclass[reprint, superscriptaddress, preprintnumbers, amsmath,amssymb, aps, prx, ]{revtex4-2}
\usepackage{graphicx, braket, float}
\usepackage[english]{babel}
\usepackage[usenames,dvipsnames,x11names]{xcolor}
\usepackage[colorlinks=true, citecolor = MidnightBlue, linkcolor = MidnightBlue, urlcolor=MidnightBlue, pdfborder={0 0 0}]{hyperref}
\usepackage{orcidlink}
\usepackage{times}
\usepackage{float}
\usepackage{newtxmath}

\begin{document}

\title{Parametric pair production of collective excitations in a {Bose}-{Einstein} condensate}

\newcommand{\LCF}{Université Paris-Saclay, Institut d’Optique Graduate School, CNRS, Laboratoire Charles Fabry, 91127, Palaiseau, France}
\author{Victor Gondret\,\orcidlink{0009-0005-8468-161X}}
\email{victor.gondret@normalesup.org}
\affiliation{\LCF}
\author{Rui Dias\,\orcidlink{0009-0004-4158-7693}}
\affiliation{\LCF}
\author{Clothilde Lamirault\,\orcidlink{0009-0001-6468-2181}}\email{clothilde.lamirault@institutoptique.fr}
\affiliation{\LCF}
\author{Léa Camier\,\orcidlink{0009-0003-4345-3608}}
\affiliation{\LCF}
\author{Amaury Micheli\,\orcidlink{0000-0002-5240-140X}}\email{amaury.micheli@riken.jp}
\affiliation{RIKEN Interdisciplinary Theoretical and Mathematical Sciences (iTHEMS), Wako, Saitama 351-0198, Japan}
\author{Charlie~Leprince\,\orcidlink{0009-0002-5490-6767}}
\affiliation{\LCF}
\author{Quentin Marolleau\,\orcidlink{0009-0002-3587-3912}}\altaffiliation{Present address: Qblox, Delftechpark, Netherlands.}
\affiliation{\LCF}
\author{Scott Robertson\,\orcidlink{0000-0001-5919-8320}}\email{scott-james.robertson@cnrs.fr}
\affiliation{Institut Pprime, CNRS -- Université de Poitiers -- ISAE-ENSMA. TSA 51124, 86073 Poitiers Cedex 9, France}
\author{Denis Boiron\,\orcidlink{0000-0002-2719-5931}} \email{denis.boiron@institutoptique.fr}\affiliation{\LCF}
\author{Christoph I. Westbrook\,\orcidlink{0000-0002-6490-0468}} \email{christopher.westbrook@institutoptique.fr}
\affiliation{\LCF}

\begin{abstract}
By exciting the transverse breathing mode of an elongated Bose-Einstein condensate, we parametrically produce longitudinal collective excitations in a pairwise manner. This process also referred to as Faraday wave generation, can be seen as an analog to cosmological particle production.  Building upon single particle detection, we investigate the early time dynamics of the exponential growth and compare our observations with a Bogoliubov description. The growth rate we observe experimentally is in very good agreement with theoretical predictions, demonstrating the validity of the Bogoliubov description and thereby confirming the smallness of quasiparticle interactions in such an elongated gas. We also discuss the presence of oscillations in the atom number, which are due to pair correlations and to the rate at which interactions are switched off.
\end{abstract}

\maketitle

\section{Introduction}

Parametric resonance is a ubiquitous phenomenon in physics, manifesting in systems ranging from nonlinear optical amplifiers to particle creation in the early Universe~\cite{kofman.1994.reheating}.
It was first reported in 1831 when Faraday  observed the spontaneous formation of surface wave patterns on a fluid subjected to vertical oscillations~\cite{faraday_xvii_1831}.
The oscillation periodically modifies the effective gravitational field and thus the dispersion relation, 
which parametrically excites modes whose frequencies are multiples of half the driving frequency~\cite{miles.1990.parametrically}.
The amplitude of these modes grows exponentially from an initial seed, for instance thermal or vacuum fluctuations, forming a pattern at the interface, before the growth saturates because of nonlinearities~\cite{edwards_patterns_1994}.

In quantum fluids such as Bose-Einstein condensates (BECs), the dispersion relation depends on the interaction strength between atoms, which can be modulated in time~\cite{staliunas.2002.faraday, shukuno_faraday_2023}, resulting in the formation of Faraday patterns~\cite{engels.2007.faraday, nicolin.2007.faraday, nguyen.2019.parametric}.
These patterns have been further interpreted as time crystals~\cite{smits.2018.observation,smits.2021.spontaneous,Smits_2020} and have revealed a wide variety of structures in two-dimensional gases~\cite{fu.2018.density, kwon.2021.spontaneous, zhang.2020.pattern, liebster.2025.prx}.
The role of vacuum fluctuations in seeding the growth of the pattern was revealed through the observation of entanglement between waves of opposite momentum~\cite{gondret.2025.observation}.
Faraday waves were also explored in other configurations, including fermionic clouds~\cite{capuzzi.2008.faraday, hernandez.2021.faraday}, two-component superfluids~\cite{cominotti.2022.observation}, and fluids with different dispersion relations, such as those exhibiting a roton minimum~\cite{lakomy.2012.faraday} or in optical lattices~\cite{capuzzi.2011.suppression, dupont.2023.emergence}.

The dynamics of these collective excitations, or quasiparticles, on top of the fluid is analogous to the dynamics of a quantum field on a curved space-time~\cite{Unruh-1981}.
Thus, by engineering an appropriate background profile, these experiments can reproduce well-known effects of quantum field theory~\cite{Barcelo2011,Schutzhold.2025.ultracoldatoms},
such as analogs of Hawking radiation~\cite{garay.2000.sonic},  cosmological particle creation~\cite{fedichev.2004.cosmological,fischer.2004.quantum, Uhlmann_2005, piyush.2007.analog}, or the dynamical Casimir effect~\cite{carusotto.2010.density, tian.2018.roton}.
However, in order to interpret or design these experiments as faithful analogs of these effects, precise modeling of the system is required, an endeavor to which R. Parentani made major contributions~\cite{macher.2009.black-hole, busch.2013.dce.dissipative, busch.2014.dce.quantum, robertson.2017.controlling, robertson.2017.assessing}.
These proposals successfully led to the experimental study of quasiparticle production in a time-dependent background~\cite{wilson.2011.observation, jaskula.2012.acoustic, Lahteenmaki:2011cwo, vezzoli.2019.optical, hu.2019.quantum, chen.2021.observation, viermann_quantum_2022, steinhauer_analogue_2022, sparn.2024.experimental} or in the presence of an analog horizon~\cite{philbin.2008.fiber, weinfurtner.2011.measurement, steinhauer.2016.observation, euve.2016.observation, munoz_de_nova_observation_2019, observation.2019.drori, svancara_rotating_2024, falque.2025.polariton}.

In this work, we use single particle detection to investigate the early-time development of Faraday patterns in a modulated BEC and give an interpretation within Bogoliubov theory, in light of Parentani’s work~\cite{busch.2014.dce.quantum, robertson.2017.controlling, robertson.2017.assessing}. In Section~\ref{sec:exp.setup},
we present the experimental setup and procedure, along with a brief summary of our model. 
In Sections~\ref{sec:result} and \ref{sec:growth}, we observe the exponential growth of the quasiparticle number and we compare the observed growth rate to a theoretical treatment~\cite{busch.2014.dce.quantum}.
Section~\ref{sec:oscillation} provides an interpretation of the oscillatory behavior of the atom number observed during the growth dynamics. This oscillation results from the presence of correlations between modes of opposite momentum, as well as from the non-adiabatic transfer between the collective state confined in the trap and the free particles which we ultimately detect.

\section{Experimental setup and model}\label{sec:exp.setup}

\begin{figure*}[tbp]
\includegraphics[width=5.51in]{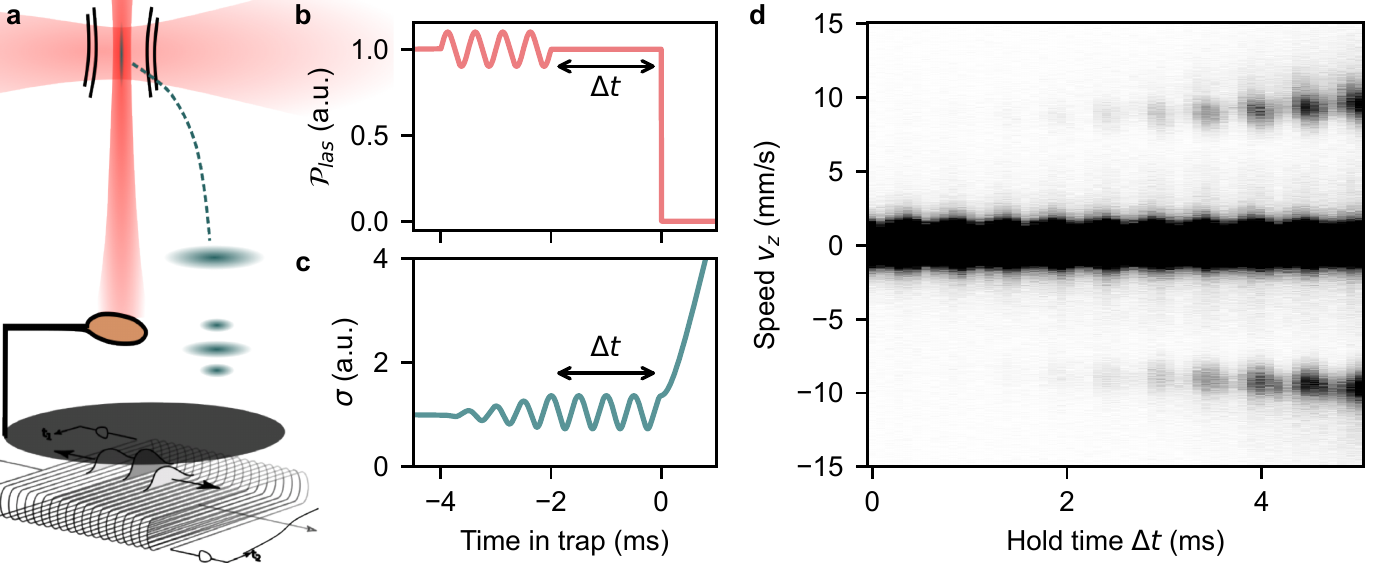}
\caption{\textbf{Experimental parametric excitation of quasiparticles.} \textbf{a}, Diagram of the experimental setup. \textbf{b}, By modulating dipole trap laser power $\mathcal{P}_{las}$, the breathing mode of the BEC is excited. \textbf{c}, The BEC width $\sigma$ oscillates as long as the cloud is kept in the trap for a duration $\Delta t$.
 The breathing mode parametrically excites two longitudinal excitations with opposite momenta. \textbf{d}, Time evolution of the longitudinal density profile of the atom velocity distribution.
 The dark band at $v_z=0$ is the condensate. After a hold time of about 2~ms, sidebands appear on either side of the condensate. 
 }
\label{fig1}
\end{figure*}

\textit{Modulation protocol --} We produce a cigar-shaped BEC with typically 15,000 atoms and at a temperature between 30 and 45~nK, in a crossed dipole trap with longitudinal and transverse frequencies of 40~Hz and 1~kHz, respectively.
The experimental setup and procedure, sketched in Fig.~\ref{fig1}a, are the same as in Ref.~\cite{gondret.2025.observation}, with the main difference being the number of atoms in the BEC and its temperature, which we control by adjusting the final trap power.
To excite the gas, the power of the vertical laser $P_{\rm las}$, which confines the gas transversely, is modulated at twice the transverse trap frequency $\omega_\perp$, for four oscillation periods, see Fig.~\ref{fig1}b.
This modulation excites the transverse breathing mode of the BEC: its width $\sigma$ continues to oscillate with the same amplitude for an additional hold time $\Delta t$ after the modulation ends~\cite{chevy.2002.transverse}, as drawn in Fig.~\ref{fig1}c (see also the experimental data in Fig.~\ref{fig3}a).
This breathing mode is not damped by thermal phonons~\cite{jackson.accidental.2002} but rather couples to longitudinal collective excitations i.e. Faraday waves~\cite{pitaevskii.1998.elementary}.

\textit{Detecting atoms --}
When the trap is switched off, the density decreases and atoms that carry a collective excitation separate from the BEC, forming two sidebands, as shown in Fig.~\ref{fig1}d.
Atoms fall onto a microchannel plate detector located 46~cm below the trap, which records the arrival time and position of single atoms with a quantum efficiency of 25(5)\%~\cite{lopes.2015.atomic}.
The time of flight is sufficiently long that the arrival time and position at the detector accurately reflect the three-dimensional velocity of the atoms at the moment the trap and interactions were switched off.
The detector is protected from the vertical trapping laser; thus the 13~\textmu s Raman pulse that transfers the atoms to a magnetically insensitive state after the trap switch off also imparts a transverse momentum, ensuring that the atoms are detected on the unshielded region of the detector.
In addition, in some data sets, a 1~ms velocity-selective Bragg pulse deflects 97\% of the BEC atoms upwards at 48~mm/s, while not affecting the sidebands~\cite{leprince.2024.coherent}.
The Bragg pulse helps prevent saturation of the detector in the vicinity of the sidebands, which may alter their shape and intensity.

\textit{Modeling quasiparticle production --}
To model the production of longitudinal collective excitations or quasiparticles, we make use of the Bogoliubov description and refer the interested reader to the appendix or to Refs.~\cite{robertson.2017.controlling, micheli.2022.phonon} for the details of the model. In this approximation, the BEC is treated as a coherent state acting as a reservoir for non-interacting quasiparticles.
For simplicity, we neglect the weak longitudinal harmonic trapping, which is not expected to change the physics at play~\cite{butera.2021.position}. 
Quasiparticles with energy $\hbar\omega_k$ and momentum $\hbar k$ are then described by annihilation (creation) operators $\hat{b}_k^{(\dagger)}$ whose evolution obeys~\cite{robertson.2017.controlling}:

\begin{equation}
\label{eq:BdG.bk}
\partial_t \hat{b}_k  = - i\omega_k\hat{b}_k + \frac{\dot{\omega}_k}{2\omega_k}  \hat{b}_{-k}^\dagger \quad \text{where} \quad \omega_k = \sqrt{ \frac{g_1 n_1}{m} k^2+\left(\frac{\hbar k^2}{2m}\right)^2}.
\end{equation}
Here, $\hbar$ is the reduced Planck constant, $m$ is the mass of the atoms, $n_1$ is the one-dimensional density and $\dot{\omega}_k$ represents the time derivative of $\omega_k$. 
The effective 1D interaction strength $g_{1}$ is proportional to the mean transverse density, and therefore inversely proportional to the cross-sectional area of the BEC.
Assuming a Gaussian transverse profile and extending the result of Refs~\cite{gerbier.quasi1d.2004, salasnich.2002.effective} to the anisotropic case, 
we find $g_1 = 2 \hbar^{2} a_s / (m \sigma_x \sigma_y)$, where $a_s$ is the atomic $s$-wave scattering length and where $x$ and $y$ correspond to the principal axes along which the transverse profile has minimal and maximal width.
During the breathing oscillation, $\sigma_{x}$ and $\sigma_{y}$ vary in time which in turn modulates the dispersion relation in Eq.~(\ref{eq:BdG.bk}).
In the absence of oscillation, $\dot{\omega}_k=0$ and the second term in Eq.~(\ref{eq:BdG.bk}) vanishes. Thus, each quasiparticle mode $\hat{b}_k$ evolves independently as $\hat{b}_k\propto e^{-i\omega_kt}$.
However, when $g_1$ varies, we have $\dot{\omega}_k\neq 0$ and the evolution of modes with opposite wavevector are coupled. 
In particular, when $\dot{\omega}_k$ oscillates at $2\omega_\perp$, the two modes with $\omega_{\pm k}=\omega_\perp$
are parametrically excited and squeezed.
The number of quasiparticles $ \langle \hat{b}_k^{\dagger} \hat{b}_k \rangle$ increases exponentially, as does their anomalous correlation $|\langle \hat{b}_k \hat{b}_{-k} \rangle|$, signaling a pair production process.
The rate of the exponential growth depends on the breathing amplitude, which itself depends on the amplitude and duration of the laser power modulation.

\textit{Mapping quasiparticles to atoms --} When the trap is switched off, the atomic wavefunction expands and the collective excitation state is mapped onto the free atomic state which is then detected.
The (a)diabatic nature of this mapping depends on the rate of change of the interaction strength $g_1n_1$~\cite{tozzo.2004.phonon} and will be discussed further in Sec.~\ref{sec:oscillation}.

\section{Exponential creation of quasiparticles}\label{sec:result}

\begin{figure}[tbp]
    \centering
    \includegraphics[width=8.6cm]{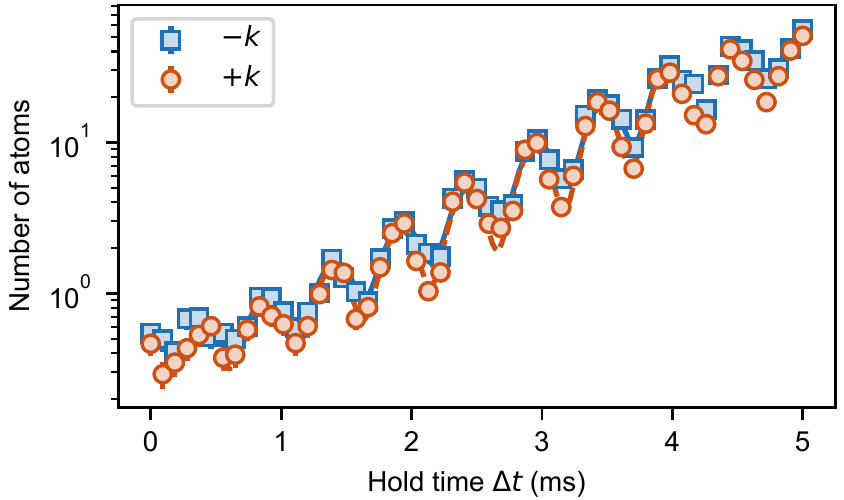}
    \caption{\textbf{Detected atom number in each sideband as a function of the hold time.} In this data set the laser power was modulated by ±15\%. The atom number was counted in each voxel as defined in the text. The data was fitted to Eq.~(\ref{eq:template_growth}) in the exponential regime, that is between 0.5 and 4~ms. The solid blue and dashed orange lines show the fits for the negative and positive velocity sidebands respectively. The fits give a reduced chi-squared of $\chi_{\nu}^{2} = 1.0$ and $\chi_{\nu}^{2} = 1.2$ respectively. }
    \label{fig2}
\end{figure}

\textit{Observations -- } Fig.~\ref{fig1}d shows the one-dimensional density in velocity space as a function of the hold time $\Delta t$ during which the BEC continues to breathe. 
In the figure, visible sidebands emerge after about 2~ms, rapidly growing in intensity and oscillating at the modulation frequency $\omega_\perp$. We will focus on these two features — growth and oscillations of the sidebands — in the following.
One also sees that the mean momentum of the sidebands increases with time, a behavior which may be related to an excitation of the longitudinal breathing mode~\cite{smits.2018.observation}. 
Lastly, we note that the longitudinal width of the BEC (at $v_z = 0$) also exhibits oscillations. This effect may be an artifact caused by detector saturation varying with the transverse width. In any case, since it affects only the BEC and not the sidebands, we consider it unimportant for our analysis.

To quantify the growth, we define a 3D volume in velocity space, a voxel, whose longitudinal size is roughly that of one mode~\cite{gondret.2025.observation}.
For each hold time $\Delta t$, each voxel is centered at the velocity for which the atom density is maximum, one on the positive velocity sideband and the other one on the negative one.
In Fig.~\ref{fig2}, the evolution of the number of atoms in the two voxels is plotted as a function of hold time, confirming the exponential growth.
The oscillations shown here have also been observed in other experiments~\cite{smits.2018.observation, hernandez.2021.faraday, liebster.2025.prx, fu.2018.density} and we will discuss their interpretation in section~\ref{sec:oscillation}.

\textit{Model --} We fit the measured number of atoms in the sidebands using

\begin{equation}
\label{eq:template_growth}
n_k^d(\Delta t) = n_0 +  \Delta n\,  e^{ G_k\Delta t }\times \left[1 + A_k  \cos \left(2\omega_k \Delta t + \varphi_k \right) \right].
\end{equation}
The model fits well to the data in Fig.~\ref{fig2}.
The physical meaning of some fit parameters, whose theoretical expression is given in the appendix, is as follows.
The offset $n_0$ accounts for other non-excited modes which are detected in the voxel analysis. 
The exponential growth is multiplied by an overall prefactor $\Delta n$, which characterizes the fluctuations that seed the parametric growth.
The parameter $G_k$ characterizes the growth and $A_k$ the amplitude of the oscillations.
In our fit, $n_0$ and $\Delta n$ are sensitive to the analysis voxel size, while $G_k$ and $A_k$ are not. These will be discussed in the next sections.

\section{Measurement of the growth rate}\label{sec:growth}

If we assume that the interaction strength $g_1$ is modulated as 
\begin{equation}\label{eq:exc}
   g_1(t)=g_{1}(0)\left[1+a \cos(\omega_d t)\right] 
\end{equation} 
where $a$ is the modulation amplitude and $\omega_d$ the driving frequency, the growth dynamics can be solved analytically. 
Without damping, the growth rate for the resonant mode $\omega_k=\omega_d/2$ in a homogeneous gas is predicted to be~\cite{busch.2014.dce.quantum}
\begin{equation}\label{eq:theoretical_growth_rate}
    G_k^{\rm th} = \frac{a}{2}\frac{\omega_{k}}{1+k^2\xi^2/4} 
\end{equation}
where  $\xi=\sqrt{\hbar^2/(m g_1n_1)}$ is the healing length of the BEC.

\begin{figure}
    \centering
    \includegraphics[width=3.375in]{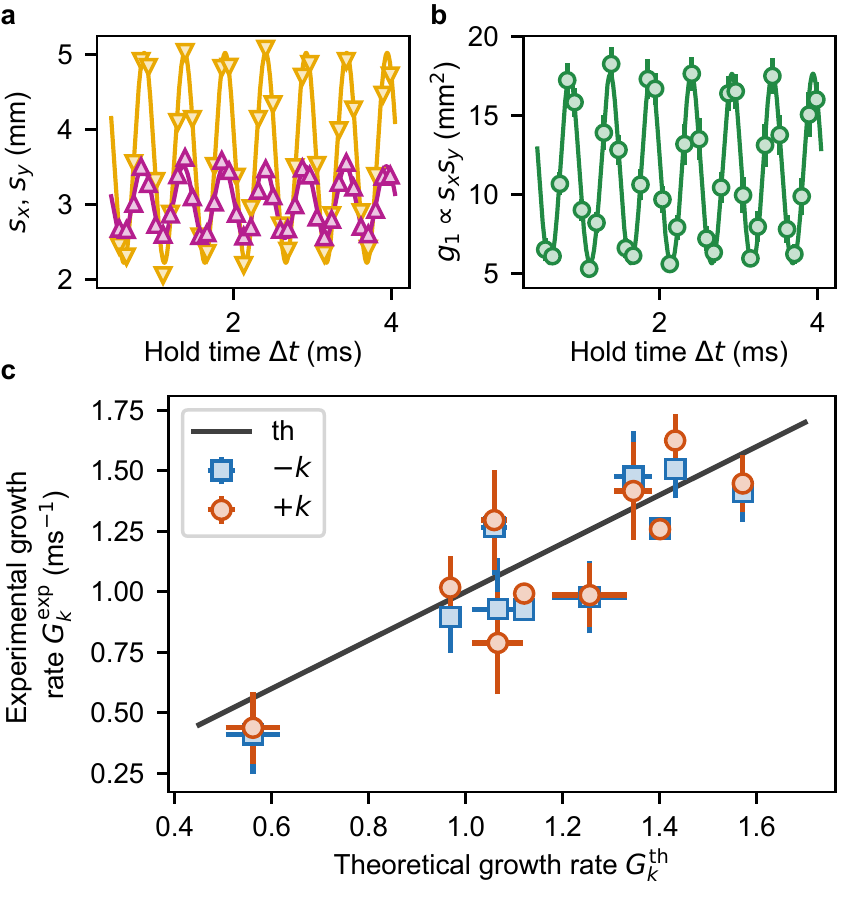}
    \caption{\textbf{Comparison between the experimental and homogeneous undamped theoretical rate.} \textbf{a}, The asymmetric breathing mode of the BEC is fitted with a sine function along both axes. The modulation parameters are the same as in Fig.~\ref{fig2}. \textbf{b}, The relative, effective 1D interaction strength. \textbf{c},
   The fitted growth rate for different modulation amplitudes on the vertical axis is compared to the prediction of Eq.~(\ref{eq:theoretical_growth_rate}) on the horizontal axis. The theoretical growth rate is also shown as a line of unit slope.
   }
    \label{fig3}
\end{figure}
\textit{Numerical estimates --} 
The healing length is estimated via the dispersion relation~\eqref{eq:BdG.bk} by matching the sidebands velocity and the trap frequency.
Since the quasiparticles are predominantly produced where the density is highest, our estimate amounts to characterizing the cloud near the center.
The effective 1D interaction strength is inversely proportional to the cross-sectional area of the BEC, $g_1\propto 1/\sigma_x \sigma_y$.
We can measure the relative amplitude of the oscillation by measuring the transverse widths $s_{x}$ and $s_y$ of the cloud at the detector for different hold times $\Delta t$.
Due to the inverse relationship between spatial confinement and rate of expansion, the width at detection is inversely proportional to the \textit{in situ} width, $s_x\propto1/\sigma_x$.
We observe a 50\% asymmetry in the breathing amplitude between the $x$ and $y$ directions, as shown in Fig.~\ref{fig3}a, which might reflect a small trap anisotropy~\footnote{We believe this anisotropy to be small because both radii breathe at the same frequency.}.
The time dependence of $s_x$ and $s_y$ is well modeled by a sine function from which we extract the amplitudes $a_x$ and $a_y$. 
Because the effective interaction strength depends on their product $g_1\propto s_xs_y$, the value of $a$ is given by $a = (a_x+a_y)/(1+a_xa_y/2)$.
The evolution of $g_1$ is shown in Fig.~\ref{fig3}b and the solid line is the product of the fits of $s_x$ and $s_y$, which shows a good agreement.

\textit{Results --} We show in Fig.~\ref{fig3}c the measured growth rate $G_k^{\rm }$ as a function of the theoretical growth rate $G_k^{\rm th}$, evaluated using Eq.~(\ref{eq:theoretical_growth_rate}). 
The line of unit slope thus corresponds to the theoretical prediction,
and the measured growth rates  are seen to be in very good  agreement with it the predicted ones.
Recently, it was shown that the growth rate should be sensitive  to interactions between quasiparticles~\cite{micheli.2022.phonon} 
arising  from higher-order terms in the Hamiltonian, which are not taken into account within the Bogoliubov description.
The dominant effect is the presence of a decay rate that acts counter to the growth of quasiparticle populations and correlations~\cite{micheli.2022.phonon, micheli.2024.dissipative}. 
With our experimental parameters, the magnitude of this decay is predicted to be on the order of 0.1~ms$^{-1}$, and while our data are not sufficiently precise to confirm such a small reduction, they do confirm that any decay rate present cannot be much larger.

\section{Microscopic interpretation and scaling of the oscillations}
\label{sec:oscillation}

The large oscillation in the exponential growth of the atom number shown in Fig.~\ref{fig2} was also observed in other experiments~\cite{hernandez.2021.faraday, liebster.2025.prx, fu.2018.density,smits.2018.observation}.
Here we analyze this oscillation within the Bogoliubov framework and show that it stems from two ingredients: the anomalous correlation between opposite momentum quasiparticles and the diabatic transfer from the interacting basis to the non-interacting one.

\textit{Bogoliubov transformation --} During the parametric excitation, the number of quasiparticles grows exponentially.
We refer here to the quasiparticle operators as the operators $\hat{b}_k$ which diagonalize the Hamiltonian at any time.
These quasiparticles correspond to superpositions of $N$ atoms moving in one direction and $N-1$ moving in the other.
Since quasiparticles are produced coherently in $\pm k$ pairs, the superposition of these moving atoms creates a standing wave in the atomic density, which oscillates at frequency $\omega_k$ as observed in Refs.~\cite{smits.2018.observation,Smits_2020,smits.2021.spontaneous, hernandez.2021.faraday}. 
This standing wave corresponds to a superposition of equal numbers of atoms at momenta $k$ and $-k$, whose number is minimal whenever the density is momentarily stationary. As this occurs twice per cycle, the atom numbers oscillate at frequency $2\omega_{k}$.  At (almost) any instant there is thus an excess of moving atoms above the minimum, and this excess can be interpreted physically as driving the movement of the density pattern, i.e. as those atoms which are moving from regions of decreasing density into regions of increasing density.

Mathematically, the quasiparticle and atomic operators $\hat{b}_k$ and $\hat{a}_k$ are related by a Bogoliubov transformation $\hat{a}_k = u_k \hat{b}_k + v_k \hat{b}_{-k}^{\dagger}$ and characterized by $(u_k,v_k)$ coefficients~\footnote{The Bogoliubov coefficients are given by $u_k, \, v_{k} = \frac{1}{2}\left( \sqrt{\hbar k^2/(2m\omega_k)}   \pm \sqrt{2\omega_km/(\hbar k^2)} \right)$.
This transformation depends on the quasiparticle frequency $\omega_k$, which itself depends on the interaction term $g_1n_1$, see Eq.~(\ref{eq:BdG.bk}).}. The number of atoms $n_k^{\rm at} = \langle \hat{a}_{k}^{\dagger} \hat{a}_{k} \rangle$ with momentum $k$ can be evaluated in the quasiparticle basis $\hat{b}_k$ and is given by 
\begin{equation}
    n_{k}^{\rm at} = v_{k}^{2} + \left(u_{k}^{2}+v_{k}^{2}\right) \langle \hat{b}_{k}^{\dagger} \hat{b}_{k} \rangle +2u_kv_k\text{Re}\left(\langle\hat{b}_{k}\hat{b}_{-k}\rangle\right)\,,
    \label{eq:atom_number_insitu}
\end{equation}
where we have used $\langle\hat{b}_k^\dagger\hat{b}_k\rangle=\langle\hat{b}_{-k}^\dagger\hat{b}_{-k}\rangle$.
When $g_1$ is modulated, all the terms of Eq.~(\ref{eq:atom_number_insitu}) experience an oscillation. 
However, the amplitudes of $u_k, v_k$ and $\langle\hat{b}_k^\dagger\hat{b}_k\rangle$ are small and the large oscillation in the atom number is caused by the last term of Eq.(\ref{eq:atom_number_insitu}), see Appendix \ref{app:modulation}.

If the number of atoms $n_k^{\rm at}$ is directly probed within the trap, 
Eq.~(\ref{eq:atom_number_insitu}) must be evaluated within the trap. In the regime of exponential growth $G_k t\geq 1$, we have $|\langle\hat{b}_k\hat{b}_{-k}\rangle|\sim \langle \hat{b}_{k}^{\dagger} \hat{b}_{k}\rangle $ and the relative oscillation amplitude $A_k$ only depends on the $(u_k,v_k)$ Bogoliubov coefficients
\begin{equation}
\label{eq:max_A}
  A_k^{\mathrm{inst}} = \frac{2 \left \lvert u_k \right \rvert  \left \lvert v_k \right \rvert}{u_k^2+v_k^2}=\left(1+\frac{\hbar^2 k^2}{2 m g_1 n_1}\right)^{-1} 
\end{equation}
which is approximately equal to 0.7 with our parameters. 
However, experiments that measure atoms in momentum space do so either after time-of-flight expansion (as in our case) or by imaging atoms following a phase-space rotation~\cite{tung.2010.observation, murthy.2014.matter}, as in~\cite{liebster.2025.prx}. In both cases, the measurement is performed once the atoms cease to interact and quasiparticles and atom operators are formally identical: $u_k=1$ and $v_k=0$ in Eq.~\eqref{eq:atom_number_insitu}. The amplitude of the observed oscillation in $n^{\rm at}_k$ then depends on the details of the measurement protocol which modifies the state of the collective excitations, e.g.~the values $\langle {b}_k^\dagger {b}_k\rangle$ and $\langle {b}_{-k}{b}_k\rangle$, while mapping it to that of the atoms. For a time-of-flight expansion, this modification depends on the rate of change of the quasiparticle frequency $\dot{\omega}_k /\omega_k$ as can be seen from Eq.~\eqref{eq:BdG.bk} where a high value of $\dot{\omega}_k /\omega_k$ induces a mixing between the $k$ and $-k$ modes, see Appendix~\ref{appendix_A2} for details. This point has also been discussed in the context of other analog gravity experiments in~\cite{alvaro.2025.relevance}.

\textit{(A)diabatic mapping --} If the interaction is switched off suddenly with respect to $1/\omega_k$, the state of the system is instantaneously projected onto the atomic plane wave basis in a way similar to an \textit{in situ} fluorescence measurement. Experimentally this situation is realized when interactions are tuned using a Feshbach resonance ~\cite{chen.2021.observation,liebster.2025.prx} or in an optical lattice \cite{dupont.2023.emergence, tenart.2021.observation}.
On the other hand, if the interaction term $g_1n_1$ decreases slowly so that $\dot{\omega}_k\ll\omega_k^2$, the system is in the so-called phonon evaporation regime~\cite{tozzo.2004.phonon}, and the collective excitation state is mapped adiabatically onto the atomic basis.

\begin{figure}
    \centering
    \includegraphics[width=3.375in]{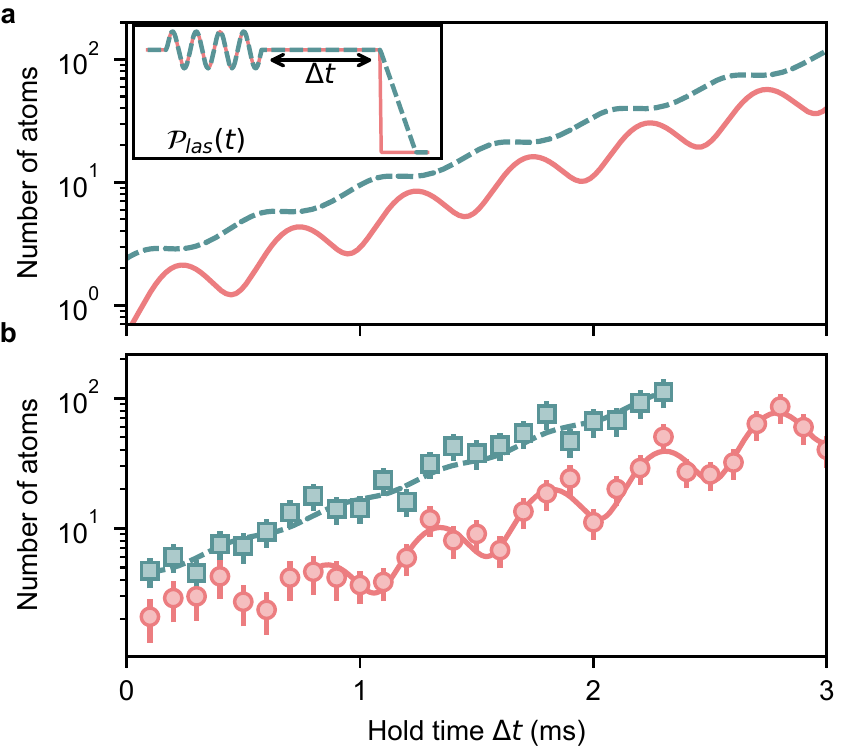}
    \caption{\textbf{A more adiabatic mapping of the quasiparticles to the atoms.} \textbf{a}, Numerical solution and \textbf{b}, experimental plot of the evolution of the number of atoms when the transverse laser power, shown in the inset, is abruptly turned off (solid and round markers, in red) or ramped down in 1.5 ms (dashed line and square markers, in green).
  }
    \label{fig4}
\end{figure}
\textit{Intermediate case --} In our experiment, the interaction strength is driven by the transverse width of the BEC $g_1\propto \sigma^{-2}$. Even if the dipole trap is turned off instantaneously, the transverse width increases on a time scale of the order of $\omega_\perp^{-1}$. As the frequency of the excited quasiparticles corresponds precisely to this timescale, we are neither in the adiabatic nor in the diabatic regime. The observed oscillation amplitudes in Figs.~\ref{fig2} and \ref{fig4}b are about 0.5, smaller than  $A_k^{\rm inst}=0.7$ and close to the amplitude found numerically (see Fig.~\ref{fig4}a). Although we cannot strictly reach it experimentally, we can approach the adiabatic case by slowly ramping down the transverse laser power.
The resulting evolution of the atom number for a 1.5~ms ramp is shown in green squares Fig.~\ref{fig4}b. One sees that the oscillations, if present, are quite weak.
On the other hand, the total number of atoms is higher than for the sudden turnoff. This is because the pair production mechanism continues to operate during the ramping down of the laser power (see Appendix~\ref{app:numerics}).

\section{Conclusion}
In this paper, we have studied the production of quasiparticles in a parametrically driven BEC.
Our observations of exponential growth and oscillation of the particle number are in agreement with Bogoliubov theory as applied in Ref.~\cite{robertson.2017.controlling}. 
 The data clearly shows the parametric nature of the process and the fact that the excitations are generated in a pairwise manner. 
 
The measured growth rate is in very good  agreement with the predicted value, even in the presence of a trap inhomogeneity. The precision of our measurements cannot isolate and measure the small effect of quasiparticle interactions, which should reduce the growth rate~\cite{micheli.2022.phonon}.
Nevertheless, our results do confirm the smallness of this decay, if present. Future work will aim to improve the experimental procedure in order to increase the signal-to-noise ratio and further isolate this effect.
Higher experimental precision may require a closer correspondence between the experiment and the model, especially if we wish to compare the value of the damping of the collective excitations with the prediction of Ref.~\cite{micheli.2022.phonon}. Thus we envision repeating the above experiments in a square potential~\cite{navon_quantum_2021}.  Alternatively, the model could be improved by including the density inhomogeneities. 
Also, exciting the gas with a Feshbach resonance would
allow to excite other $k$ modes while our excitation method is restricted to modes satisfying $\omega_k=\omega_\perp$~\footnote{By varying the excitation frequency of the laser power, we could in principle excite other modes~\cite{jaskula.2012.acoustic}. However, the procedure did not yield such a clear parametric growth as in Fig.~\ref{fig2}.}.
This perspective is interesting because the predicted decay rate of Ref.~\cite{micheli.2022.phonon} depends on~$k$.

The observed oscillations are well understood. If we could precisely measure their amplitude and estimate $u_k$ and $v_k$, we would be able to compare the quasiparticle population to their anomalous correlation (see Eq.~(\ref{eq:atom_number_insitu})).
This comparison would demonstrate the (non)separability of the two-mode state~\cite{robertson.2017.assessing} without looking at many-body correlation functions~\cite{gondret.2025.quantifying}.

\section*{Acknowledgments}
We thank and dedicate this work to Renaud Parentani, to whom we owe the genesis of this study. We acknowledge funding from QuantERA Grant No. ANR-22-QUA2-000801 (MENTA) and ANR Grant No. 20-CE-47-0001-01 (COSQUA), the LabEx PALM (ANR-10-LABX-0039PALM). 
V.G. was funded by the Région Ile-de-France in the framework of the DIM SIRTEQ program and the Quantum Saclay program FQPS (ANR-21-CMAQ-0002). 
R.D. acknowledges a Ph.D grant with reference 2024.03181.BD from the Portuguese Foundation for Science and Technology (FCT).
S.R. is funded by the CNRS Chair in Physical Hydrodynamics (ANR-22-CPJ2-0039-01).
\appendix

\section{Model}
\label{sec:model}
In this appendix, we describe with a bit more details than in the main text our modeling  of the longitudinal excitations produced in the experiment and refer the reader to Refs.~\cite{robertson.2017.controlling,micheli.2024.dissipative} for more details.

\textit{Collective excitations --} Our gas is neither in the 3D cigar-shaped regime nor in the 1D mean-field regime~\cite{menotti.2002.collective} so we model its transverse profile using an effective Gaussian ansatz, assumed cylindrically symmetric for simplicity, characterized by its width $\sigma$~\cite{gerbier.quasi1d.2004, salasnich.2002.effective}. The evolution of $\sigma$ is controlled by the trapping frequency~\cite{robertson.2017.controlling}
\begin{equation}
\label{eq:EOM_sigma}
  \ddot{\sigma} + \omega_{\perp}^2 \left( t \right) \sigma = \frac{\hbar^2}{m^2 \sigma^3} \, .
\end{equation}
Neglecting transverse excitations, we integrate over this profile to have an effective one-dimensional gas with a contact interaction $g_1 = 2 \hbar a_s^2 / (m \sigma^2)$~\footnote{Even without assuming a Gaussian profile for the transverse density, a similar result can be obtained~\cite{robertson.2017.controlling}.}. 
The small excitations around the BEC of our weakly interacting gas are well described within Bogoliubov theory~\cite{bogoliubov.1947.theory}.  
We split the atomic field $\hat{\Psi}(z)$ into a classical piece $\Psi_0$ describing the homogeneous BEC, and a quantum perturbation $\delta \hat{\Psi}(z)$ describing the uncondensed atoms, and truncate the Hamiltonian at second order in this perturbation $H^{(2)}$~\cite{pitaevskii.2016.bose_einstein}~\footnote{The difficulties with Bogoliubov treatment for a 1D homogenous gas~\cite{castinmora} are irrelevant here since we focus on a pair of modes only.}. The dynamics of these interacting uncondensed atoms can be conveniently recast as that of freely evolving \textit{collective} excitations, or quasiparticles. 
They are represented by annihilation (creation) operators $\hat{b}_k^{(\dagger)}$ related to the atomic one $\hat{a}_k^{(\dagger)}$ by a Bogoliubov transformation $\hat{a}_k = u_k \hat{b}_k + v_k \hat{b}_{-k}^{\dagger}$, where the values of $u_k, v_k$ are given in the main text. The quasiparticle operators by definition diagonalize the Hamiltonian $H^{(2)} = \sum_{k \neq 0} \hbar \omega_k(g_1) \hat{b}_k^{\dagger} \hat{b}_k$ with $\omega_k$ given by Eq.~(\ref{eq:BdG.bk}) and $n_1 = \lvert \Psi_0 \rvert^2$ the density of the BEC. Thus, a collective excitation of momentum $k$ has an energy $\hbar \omega_k$ and consists of atoms with opposite momenta $\pm k$ that keep interacting. 
Our experimental set-up allows us to measure atom numbers $n^{\rm at}_k = \langle\hat{a}_k^{\dagger} \hat{a}_k\rangle$ as a function of momentum $k$ which is related to the number of quasiparticles $n_{k} = \langle \hat{b}_k^{\dagger} \hat{b}_k \rangle$ and their anomalous correlations $c_{k} = \langle \hat{b}_k \hat{b}_{-k} \rangle$ by Eq.~(\ref{eq:atom_number_insitu}).\\

\textit{Dynamics --} 
Working in the Heisenberg picture, the evolution of the quasiparticle content is given by Eq.~\eqref{eq:BdG.bk} whose solution between two times $t\geq t^{\prime}$ is given by a Bogoliubov transformation on the quasiparticle operators~\cite{busch.2014.dce.quantum} 
\begin{equation}
\label{eq:btprime_to_bt}
\hat{b}_k \left( t \right) = \alpha \left( t ; t^{\prime} \right)  \hat{b}_k \left( t^{\prime} \right) + \beta^{\star} \left( t ; t^{\prime} \right) \hat{b}^{\dagger}_{-k} \left( t^{\prime} \right) \, ,
\end{equation}
where $\alpha \left( t ; t^{\prime} \right) , \beta\left( t ; t^{\prime} \right)$ are solutions of Eq.~\eqref{eq:BdG.bk} with initial conditions  $\alpha \left( t^{\prime} ; t^{\prime} \right) = 1, \beta \left( t^{\prime} ; t^{\prime} \right) =0$.
At the start of the experiment the trap is held constant $\omega_{\perp} = \omega_{\perp,0}$, the gas's transverse size $\sigma$ is fixed and so $\omega_k$ is time-independent, which leaves only the first term in Eq.~\eqref{eq:BdG.bk} encoding the free evolution $\hat{b}_k(t)=\hat{b}_{k}(0)e^{-i\omega_k t}$. The quasiparticle number is time-independent $n_k(t) = n_k(0)$ and the correlation oscillates $c_k(t) = e^{-i 2 \omega_k t} c_k(0)$ at $2 \omega_k$. 
When the trap frequency $\omega_{\perp}$ is varied, the transverse size of the gas reacts according to Eq.~\eqref{eq:EOM_sigma} and $\omega_k$ becomes time-dependent. The second term in Eq.(\ref{eq:BdG.bk}) then does lead to a production of pairs of opposite momentum quasiparticles which can be separated in the following two stages~\cite{robertson.2017.controlling}.

\subsection{Modulation}\label{app:modulation}

We initially modulate the trap frequency at resonance $2\omega_{\perp,0}$ for $\delta t =  4 \times  \pi / \omega_{\perp,0}$ after which we hold the trap at $\omega_{\perp,0}$ for a duration $\Delta t$. 
Eq.~\eqref{eq:EOM_sigma} then predicts that $\sigma$ first experiences breathing oscillations at frequency $2 \omega_{\perp,0}$ with linearly growing amplitude, which then persist at fixed amplitude during the hold time, see Fig.~\ref{fig1}.
This variation leads $g_1$ to oscillate at $2 \omega_{\perp,0}$~\cite{robertson.2017.controlling}. To get explicit expressions, we neglect the growth phase and assume sinusoidal oscillation with a fixed amplitude $a$ i.e. $g_1\left(t\right)$ is given by Eq.~(\ref{eq:exc}) from $t=0$ to $\Delta t$. 
This modulation corresponds to a  two-mode squeezing operation on the $\pm k$ modes which are resonant with the process $\omega_{\pm k}=\omega_{\perp,0}$~\cite{busch.2014.dce.quantum}. 
Assuming an initial thermal population  $n_{\pm k} (0) = n_k^{\mathrm{in}}$ and $c_k (0) = 0$, Eq.~\eqref{eq:btprime_to_bt} can be solved using a rotating wave approximation and gives~\cite{busch.2014.dce.quantum}
\begin{align}
\begin{split}
n_k \left(t\right) + \frac{1}{2} & = \left( n_k^{\mathrm{in}} + \frac{1}{2} \right) \cosh \left( G_k t \right) \, , \\
c_k \left(t\right) & = e^{- 2 i \int_0^{t} \omega_k (t^{\prime}) \mathrm{d} t^{\prime} } \left( n_k^{\mathrm{in}} + \frac{1}{2} \right)  \sinh \left( G_k t \right)  \, ,
\label{eq:n_and_c_modulation}
\end{split}
\end{align}
where the growth rate is given by Eq.~(\ref{eq:theoretical_growth_rate}).
Neglecting the early linear phase~\cite{micheli.2024.dissipative}, these quantities grow exponentially 
\begin{align}\label{eq:app.expogrowh}
    \begin{split}
        n_{k}(t) + \frac{1}{2} & \to \left(n_{k}^{\rm in}+\frac{1}{2}\right) \frac{1}{2} e^{G_{k}t} \, , \\
        c_{k}(t) & \to e^{-2i\int_{0}^{t}\omega_{k}\left(t^{\prime}\right) {\rm d}t^{\prime}} \left(n_{k}^{\rm in}+\frac{1}{2}\right) \frac{1}{2} e^{G_{k}t} \,.
    \end{split}
\end{align}

\begin{figure}
    \centering
    \includegraphics[width=\linewidth]{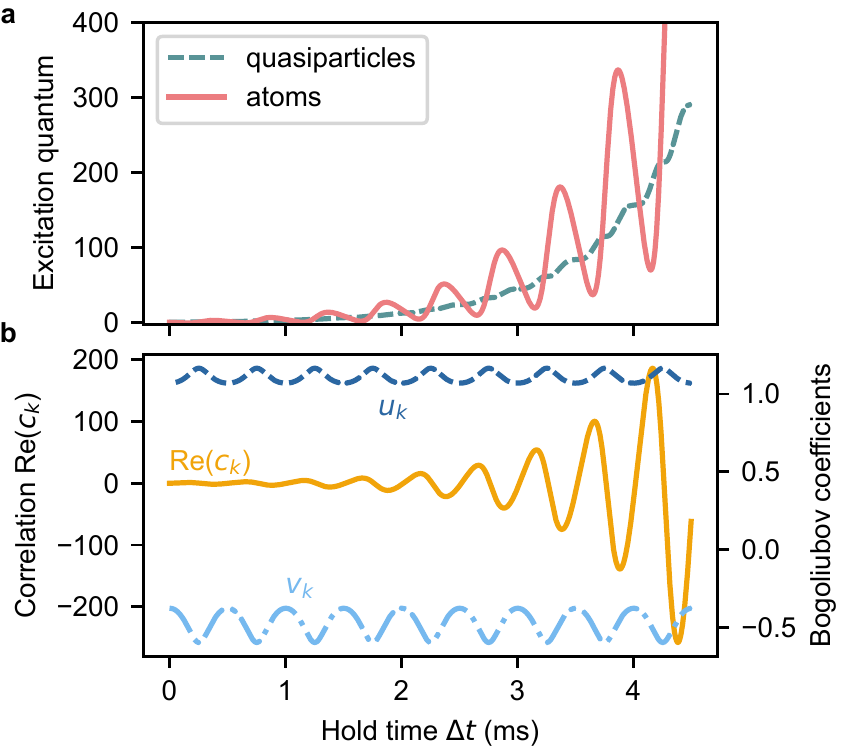}
    \caption{\textbf{Theoretical evolution of the mean value of some operators during the parametric growth. }\textbf{a}, Quasiparticle and atomic populations during the excitation.
    \textbf{b}, Corresponding evolution of the real part of the anomalous correlation $\mathrm{Re}(c_k)$ (left $y$ scale, solid yellow)  and of the Bogoliubov coefficients $u_k$ and $v_k$ (right $y$ scale, in dashed blue). 
    }
    \label{figAto}
\end{figure}

To check that Eqs~\eqref{eq:app.expogrowh} capture the essential features of the quasiparticle production process, we compare them to a more complete numerical description of the experiment, see Sec.~\ref{app:numerics} for details. The results are shown Fig~\ref{figAto}.

The number of atoms $n_k^{\rm at}$, defined in Eq.~\eqref{eq:atom_number_insitu} and shown in solid red, exhibits pronounced oscillations as witnessed experimentally. In contrast, and in agreement with Eq.~\eqref{eq:app.expogrowh}, the number of quasiparticles $n_k$, shown in dashed blue, grows exponentially with only shallow oscillations not captured by Eq.~\eqref{eq:app.expogrowh} because of the rotating wave approximation. 
Since these cannot be responsible for the oscillations of $n_k^{\rm at}$, we examine in Fig.~\ref{figAto}b, the evolution of the other terms in Eq.~(\ref{eq:atom_number_insitu}): the Bogoliubov coefficients $u_k,v_k$, shown in dashed blue lines, and the real part of the anomalous correlation $\text{Re}(c_k)$, in solid yellow. While the Bogoliubov coefficients do exhibit small oscillations, their amplitude is much smaller than that of $\text{Re}(c_k)$, which is thus the source of the large oscillations in $n_k^{\rm at}$ as claimed in the main text. Note that a non-zero value of $v_k$ in Eq.~\eqref{eq:atom_number_insitu} is also required.

\subsection{Expansion}\label{appendix_A2}

After $\Delta t$ the trap is switched off, i.e. $\omega_\perp^{2} \to 0$, the gas expands $\sigma \to \infty$, and the atoms stop interacting $g_1 \to 0$, see Fig.~\ref{fig1}. This change of interaction also affects the quasiparticle state via a Bogoliubov transform given by Eq.~\eqref{eq:btprime_to_bt} which has to be solved from opening time $t = \Delta t$ to detection time, effectively $t = \infty$ using the variation of $\sigma$ computed from Eq.~\eqref{eq:EOM_sigma}. We make this dependence implicit and write $\alpha_{\mathrm{op.}} = \alpha({\infty;\Delta t}), \beta_{\mathrm{op.}} = \beta( \infty;\Delta t )$. Since at $t = \infty$ the gas is non-interacting ($g_1=0$), we have $u_k ( \infty ) =1, v_k ( \infty )=0$ and quasiparticles correspond to atoms, in particular $n_k^{\rm at} (\infty) = n_k (\infty)$. 
The number of atoms with momentum $k$ at the end of the expansion is thus related to the quasiparticle state at the opening time via
\begin{equation}
\begin{split}
    n_k^{\rm at} \left(\infty\right)  + \frac{1}{2} =& \left( \lvert \alpha_{\mathrm{op.}}\rvert^2 + \lvert \beta_{\mathrm{op.}}\rvert^2 \right) \left[  n_k \left(\Delta t\right) + \frac{1}{2} \right] \\&+ 2 \mathrm{Re} \left[ \alpha_{\mathrm{op.}} \beta_{\mathrm{op.}}^{\star} c_k \left(\Delta t\right) \right] \, .
\end{split}
\label{eq:final_atom_number}
\end{equation}
Eq.~\eqref{eq:final_atom_number} demonstrates that the details of the opening procedure must be taken into account since they condition the way the quasiparticle number and correlations are revealed in the detected atom number~\cite{tozzo.2004.phonon}. First, the coefficients depend on the opening time $\Delta t$ that selects initial values of $\sigma$ and $\dot{\sigma}$. Yet, since $\sigma$ oscillates at fixed amplitude we expect that dependence to be weak. Second, they will depend strongly on how quickly the trap is turned off~\cite{robertson.2017.controlling}, which gives different evolutions for $\sigma$ computed from Eq.~\eqref{eq:EOM_sigma}. 

In the regime of exponential growth where $\lvert c_{k}\left(\Delta t\right) \rvert \sim n_{k}\left(\Delta t\right)$, we can immediately read off the relative amplitude of the oscillations:
\begin{equation}
    A_{k} = \frac{2 \lvert \alpha_{\rm op.} \rvert \lvert \beta_{\rm op.} \rvert}{\lvert \alpha_{\rm op.} \rvert^{2} + \lvert \beta_{\rm op.} \rvert^{2}} \approx 2 \lvert \beta_{\rm op.} \rvert \,.
\end{equation}
This is the analogue of Eq.~(\ref{eq:max_A}) when the interaction switch-off is not instantaneous, while the approximation of $2\lvert\beta_{\rm op.}\rvert$ applies when $\lvert \beta_{\rm op.} \rvert^{2} \ll \vert \alpha_{\rm op.} \rvert^{2} \approx 1$.  The relative amplitude of the oscillations thus provides us with a measurement of the strength of the Bogoliubov transformation associated with the expansion.

\subsection{Analytical growth of the atom number}
We now combine the effect of the modulation and the expansion to derive the expected growth in the atom number.

We assume that $g_1$ is modulated sinusoidally as Eq.~(\ref{eq:exc})
for a duration $\Delta t$ before opening the trap.
The number and correlation of quasiparticles are then given by Eq.~\eqref{eq:n_and_c_modulation} at opening time. 
The detected number of atoms $n_k^{\rm det}$ in the resonant mode $\omega_k=\omega_d/2$ after trap expansion is obtained by inserting these function in Eq.~\eqref{eq:final_atom_number} and multiplying the overall atom number by $\eta$ the quantum efficiency of the detector. 
We then obtain 
\begin{equation}
\begin{split}
    n_k^{\rm det}(\Delta t) =& n_0 + 2 \cosh \left(  G_k\Delta t \right) \times \, \Delta n\\ & \times \left[1 + A_k \tanh \left( G_k \Delta t \right)  \cos \left(2\omega_k \Delta t + \varphi_k \right) \right] .
\end{split}
\end{equation}
with 
\begin{align}
\begin{split}
n_0 & = - \frac{\eta}{2} \, , \\
\Delta n & = 
\eta \left( n_k^{\mathrm{in}} + \frac{1}{2}  \right) \frac{1}{2} \left( \lvert \alpha_{\mathrm{op.}} \rvert^2 + \lvert \beta_{\rm op.} \rvert^{2} \right)  \, , \\
A_k & = \frac{2 \lvert \alpha_{\mathrm{op.}} \rvert \lvert \beta_{\mathrm{op.}} \rvert }{\lvert \alpha_{\mathrm{op.}} \rvert^2 + \lvert \beta_{\mathrm{op.}} \rvert^2 } 
\, , \\
\varphi_k  & = 2 \int \left[ \omega_{\perp} - \omega_k \left(t \right) \right] - \mathrm{Arg} \left[ \alpha_{\mathrm{op.}} \beta_{\mathrm{op.}}^{\star} \right] \, ,
\end{split}
\label{eq:many_equations}
\end{align}
which for $G_k\Delta t \geq 1$ gives Eq.(\ref{eq:template_growth}).
In Eq.~(\ref{eq:template_growth}), we have suppressed the time-dependence of $\varphi$ which is expected to be small compared to the rotating phase $2 \omega_{\perp} \Delta t$.

\subsection{Numerical simulation of both phenomena}\label{app:numerics}

The predicted amplitude in Eq.~(\ref{eq:many_equations}) relies on the value of $\lvert \beta_{\rm op} \rvert$, which characterizes the Bogoliubov transformation from the quasiparticle to the atomic basis.
We can then simulate numerically the expected amplitude within our model.
Using our Gaussian ansatz, we infer $g_1 n_1$ and using the dispersion relation we find the resonant modes $k$ such that $\omega_k = \omega_{\perp}$. 
We then numerically solve the full dynamics of $\sigma$ given by Eq.~\eqref{eq:EOM_sigma}: an initial four periods of modulation at $2\omega_{\perp,0}$, a hold time of $\Delta t$, and then the opening. Using this variation we numerically solve Eq.~\eqref{eq:btprime_to_bt} from an initial thermal state of quasiparticles at temperature $T=35$~nK, to a late enough time after the trap switch-off such that the Bogoliubov coefficients have become time-independent. This gives us access numerically to the number of detected atoms Eq.~\eqref{eq:template_growth}, with a 25\% quantum efficiency. We repeat this procedure for two different switch-off ramps, a very short one of 10~\textmu s and a longer one of 1.5 ms, matching that of the experimental realizations shown in Fig.~\ref{fig4}b. The corresponding evolution of the atom number is shown in Fig.~\ref{fig4}a. 
The solid red line corresponds to the case where the dipole trap power is switched off abruptly and the dashed green curve assumes that the dipole trap power is ramped down in 1.5~ms.
In agreement with the experimental data, we observe pronounced oscillations of relative amplitude $A_k \approx 0.34$ for the shortest ramp, roughly half that of the \textit{in situ} oscillations for which Eq.~\eqref{eq:max_A} gives $A_k^{\mathrm{inst}}\approx 0.7$, and much reduced oscillation for the slowest turn off, with an amplitude of 0.1.
As for the experimental data, we observe that the green dashed curve is above the solid red one, an indication of the additional quasiparticle production that occurs during the slow switch-off.

\bibliography{samplebib}

\begin{thebibliography}{81}%
\makeatletter
\providecommand \@ifxundefined [1]{%
 \@ifx{#1\undefined}
}%
\providecommand \@ifnum [1]{%
 \ifnum #1\expandafter \@firstoftwo
 \else \expandafter \@secondoftwo
 \fi
}%
\providecommand \@ifx [1]{%
 \ifx #1\expandafter \@firstoftwo
 \else \expandafter \@secondoftwo
 \fi
}%
\providecommand \natexlab [1]{#1}%
\providecommand \enquote  [1]{``#1''}%
\providecommand \bibnamefont  [1]{#1}%
\providecommand \bibfnamefont [1]{#1}%
\providecommand \citenamefont [1]{#1}%
\providecommand \href@noop [0]{\@secondoftwo}%
\providecommand \href [0]{\begingroup \@sanitize@url \@href}%
\providecommand \@href[1]{\@@startlink{#1}\@@href}%
\providecommand \@@href[1]{\endgroup#1\@@endlink}%
\providecommand \@sanitize@url [0]{\catcode `\\12\catcode `\$12\catcode `\&12\catcode `\#12\catcode `\^12\catcode `\_12\catcode `\%12\relax}%
\providecommand \@@startlink[1]{}%
\providecommand \@@endlink[0]{}%
\providecommand \url  [0]{\begingroup\@sanitize@url \@url }%
\providecommand \@url [1]{\endgroup\@href {#1}{\urlprefix }}%
\providecommand \urlprefix  [0]{URL }%
\providecommand \Eprint [0]{\href }%
\providecommand \doibase [0]{https://doi.org/}%
\providecommand \selectlanguage [0]{\@gobble}%
\providecommand \bibinfo  [0]{\@secondoftwo}%
\providecommand \bibfield  [0]{\@secondoftwo}%
\providecommand \translation [1]{[#1]}%
\providecommand \BibitemOpen [0]{}%
\providecommand \bibitemStop [0]{}%
\providecommand \bibitemNoStop [0]{.\EOS\space}%
\providecommand \EOS [0]{\spacefactor3000\relax}%
\providecommand \BibitemShut  [1]{\csname bibitem#1\endcsname}%
\let\auto@bib@innerbib\@empty
\bibitem [{\citenamefont {Kofman}\ \emph {et~al.}(1994)\citenamefont {Kofman}, \citenamefont {Linde},\ and\ \citenamefont {Starobinsky}}]{kofman.1994.reheating}%
  \BibitemOpen
  \bibfield  {author} {\bibinfo {author} {\bibfnamefont {L.}~\bibnamefont {Kofman}}, \bibinfo {author} {\bibfnamefont {A.}~\bibnamefont {Linde}},\ and\ \bibinfo {author} {\bibfnamefont {A.~A.}\ \bibnamefont {Starobinsky}},\ }\bibfield  {title} {\bibinfo {title} {Reheating after inflation},\ }\href {https://doi.org/10.1103/PhysRevLett.73.3195} {\bibfield  {journal} {\bibinfo  {journal} {Phys. Rev. Lett.}\ }\textbf {\bibinfo {volume} {73}},\ \bibinfo {pages} {3195} (\bibinfo {year} {1994})}\BibitemShut {NoStop}%
\bibitem [{\citenamefont {Faraday}(1831)}]{faraday_xvii_1831}%
  \BibitemOpen
  \bibfield  {author} {\bibinfo {author} {\bibfnamefont {M.}~\bibnamefont {Faraday}},\ }\bibfield  {title} {\bibinfo {title} {{On} a peculiar class of acoustical figures; and on certain forms assumed by groups of particles upon vibrating elastic surfaces},\ }\href {https://doi.org/10.1098/rstl.1831.0018} {\bibfield  {journal} {\bibinfo  {journal} {Philosophical Transactions of the Royal Society of London}\ }\textbf {\bibinfo {volume} {121}},\ \bibinfo {pages} {299} (\bibinfo {year} {1831})}\BibitemShut {NoStop}%
\bibitem [{\citenamefont {Miles}\ and\ \citenamefont {Henderson}(1990)}]{miles.1990.parametrically}%
  \BibitemOpen
  \bibfield  {author} {\bibinfo {author} {\bibfnamefont {J.}~\bibnamefont {Miles}}\ and\ \bibinfo {author} {\bibfnamefont {D.}~\bibnamefont {Henderson}},\ }\bibfield  {title} {\bibinfo {title} {Parametrically forced surface waves},\ }\href {https://doi.org/https://doi.org/10.1146/annurev.fl.22.010190.001043} {\bibfield  {journal} {\bibinfo  {journal} {Annual Review of Fluid Mechanics}\ }\textbf {\bibinfo {volume} {22}},\ \bibinfo {pages} {143} (\bibinfo {year} {1990})}\BibitemShut {NoStop}%
\bibitem [{\citenamefont {Edwards}\ and\ \citenamefont {Fauve}(1994)}]{edwards_patterns_1994}%
  \BibitemOpen
  \bibfield  {author} {\bibinfo {author} {\bibfnamefont {W.~S.}\ \bibnamefont {Edwards}}\ and\ \bibinfo {author} {\bibfnamefont {S.}~\bibnamefont {Fauve}},\ }\bibfield  {title} {\bibinfo {title} {Patterns and quasi-patterns in the {Faraday} experiment},\ }\href {https://doi.org/10.1017/S0022112094003642} {\bibfield  {journal} {\bibinfo  {journal} {Journal of Fluid Mechanics}\ }\textbf {\bibinfo {volume} {278}},\ \bibinfo {pages} {123} (\bibinfo {year} {1994})}\BibitemShut {NoStop}%
\bibitem [{\citenamefont {Staliunas}\ \emph {et~al.}(2002)\citenamefont {Staliunas}, \citenamefont {Longhi},\ and\ \citenamefont {de~Valc\'arcel}}]{staliunas.2002.faraday}%
  \BibitemOpen
  \bibfield  {author} {\bibinfo {author} {\bibfnamefont {K.}~\bibnamefont {Staliunas}}, \bibinfo {author} {\bibfnamefont {S.}~\bibnamefont {Longhi}},\ and\ \bibinfo {author} {\bibfnamefont {G.~J.}\ \bibnamefont {de~Valc\'arcel}},\ }\bibfield  {title} {\bibinfo {title} {{Faraday} patterns in {Bose}-{Einstein} condensates},\ }\href {https://doi.org/10.1103/PhysRevLett.89.210406} {\bibfield  {journal} {\bibinfo  {journal} {Phys. Rev. Lett.}\ }\textbf {\bibinfo {volume} {89}},\ \bibinfo {pages} {210406} (\bibinfo {year} {2002})}\BibitemShut {NoStop}%
\bibitem [{\citenamefont {Shukuno}\ \emph {et~al.}(2023)\citenamefont {Shukuno}, \citenamefont {Sano},\ and\ \citenamefont {Tsubota}}]{shukuno_faraday_2023}%
  \BibitemOpen
  \bibfield  {author} {\bibinfo {author} {\bibfnamefont {N.}~\bibnamefont {Shukuno}}, \bibinfo {author} {\bibfnamefont {Y.}~\bibnamefont {Sano}},\ and\ \bibinfo {author} {\bibfnamefont {M.}~\bibnamefont {Tsubota}},\ }\bibfield  {title} {\bibinfo {title} {{Faraday} {Waves} in {Bose}-{Einstein} {Condensates} -- {The} {Excitation} by the {Modulation} of the {Interaction} and the {Potential}},\ }\href {https://doi.org/10.7566/JPSJ.92.064602} {\bibfield  {journal} {\bibinfo  {journal} {Journal of the Physical Society of Japan}\ }\textbf {\bibinfo {volume} {92}},\ \bibinfo {pages} {064602} (\bibinfo {year} {2023})}\BibitemShut {NoStop}%
\bibitem [{\citenamefont {Engels}\ \emph {et~al.}(2007)\citenamefont {Engels}, \citenamefont {Atherton},\ and\ \citenamefont {Hoefer}}]{engels.2007.faraday}%
  \BibitemOpen
  \bibfield  {author} {\bibinfo {author} {\bibfnamefont {P.}~\bibnamefont {Engels}}, \bibinfo {author} {\bibfnamefont {C.}~\bibnamefont {Atherton}},\ and\ \bibinfo {author} {\bibfnamefont {M.~A.}\ \bibnamefont {Hoefer}},\ }\bibfield  {title} {\bibinfo {title} {Observation of {Faraday} waves in a {Bose}-{Einstein} condensate},\ }\href {https://doi.org/10.1103/PhysRevLett.98.095301} {\bibfield  {journal} {\bibinfo  {journal} {Phys. Rev. Lett.}\ }\textbf {\bibinfo {volume} {98}},\ \bibinfo {pages} {095301} (\bibinfo {year} {2007})}\BibitemShut {NoStop}%
\bibitem [{\citenamefont {Nicolin}\ \emph {et~al.}(2007)\citenamefont {Nicolin}, \citenamefont {Carretero-Gonz\'alez},\ and\ \citenamefont {Kevrekidis}}]{nicolin.2007.faraday}%
  \BibitemOpen
  \bibfield  {author} {\bibinfo {author} {\bibfnamefont {A.~I.}\ \bibnamefont {Nicolin}}, \bibinfo {author} {\bibfnamefont {R.}~\bibnamefont {Carretero-Gonz\'alez}},\ and\ \bibinfo {author} {\bibfnamefont {P.~G.}\ \bibnamefont {Kevrekidis}},\ }\bibfield  {title} {\bibinfo {title} {{Faraday} waves in {Bose}-{Einstein} condensates},\ }\href {https://doi.org/10.1103/PhysRevA.76.063609} {\bibfield  {journal} {\bibinfo  {journal} {Phys. Rev. A}\ }\textbf {\bibinfo {volume} {76}},\ \bibinfo {pages} {063609} (\bibinfo {year} {2007})}\BibitemShut {NoStop}%
\bibitem [{\citenamefont {Nguyen}\ \emph {et~al.}(2019)\citenamefont {Nguyen}, \citenamefont {Tsatsos}, \citenamefont {Luo}, \citenamefont {Lode}, \citenamefont {Telles}, \citenamefont {Bagnato},\ and\ \citenamefont {Hulet}}]{nguyen.2019.parametric}%
  \BibitemOpen
  \bibfield  {author} {\bibinfo {author} {\bibfnamefont {J.~H.~V.}\ \bibnamefont {Nguyen}}, \bibinfo {author} {\bibfnamefont {M.~C.}\ \bibnamefont {Tsatsos}}, \bibinfo {author} {\bibfnamefont {D.}~\bibnamefont {Luo}}, \bibinfo {author} {\bibfnamefont {A.~U.~J.}\ \bibnamefont {Lode}}, \bibinfo {author} {\bibfnamefont {G.~D.}\ \bibnamefont {Telles}}, \bibinfo {author} {\bibfnamefont {V.~S.}\ \bibnamefont {Bagnato}},\ and\ \bibinfo {author} {\bibfnamefont {R.~G.}\ \bibnamefont {Hulet}},\ }\bibfield  {title} {\bibinfo {title} {Parametric excitation of a {Bose}-{Einstein} condensate: From {Faraday} waves to granulation},\ }\href {https://doi.org/10.1103/PhysRevX.9.011052} {\bibfield  {journal} {\bibinfo  {journal} {Phys. Rev. X}\ }\textbf {\bibinfo {volume} {9}},\ \bibinfo {pages} {011052} (\bibinfo {year} {2019})}\BibitemShut {NoStop}%
\bibitem [{\citenamefont {Smits}\ \emph {et~al.}(2018)\citenamefont {Smits}, \citenamefont {Liao}, \citenamefont {Stoof},\ and\ \citenamefont {van~der Straten}}]{smits.2018.observation}%
  \BibitemOpen
  \bibfield  {author} {\bibinfo {author} {\bibfnamefont {J.}~\bibnamefont {Smits}}, \bibinfo {author} {\bibfnamefont {L.}~\bibnamefont {Liao}}, \bibinfo {author} {\bibfnamefont {H.~T.~C.}\ \bibnamefont {Stoof}},\ and\ \bibinfo {author} {\bibfnamefont {P.}~\bibnamefont {van~der Straten}},\ }\bibfield  {title} {\bibinfo {title} {Observation of a space-time crystal in a superfluid quantum gas},\ }\href {https://doi.org/10.1103/PhysRevLett.121.185301} {\bibfield  {journal} {\bibinfo  {journal} {Phys. Rev. Lett.}\ }\textbf {\bibinfo {volume} {121}},\ \bibinfo {pages} {185301} (\bibinfo {year} {2018})}\BibitemShut {NoStop}%
\bibitem [{\citenamefont {Smits}\ \emph {et~al.}(2021)\citenamefont {Smits}, \citenamefont {Stoof},\ and\ \citenamefont {van~der Straten}}]{smits.2021.spontaneous}%
  \BibitemOpen
  \bibfield  {author} {\bibinfo {author} {\bibfnamefont {J.}~\bibnamefont {Smits}}, \bibinfo {author} {\bibfnamefont {H.~T.~C.}\ \bibnamefont {Stoof}},\ and\ \bibinfo {author} {\bibfnamefont {P.}~\bibnamefont {van~der Straten}},\ }\bibfield  {title} {\bibinfo {title} {Spontaneous breaking of a discrete time-translation symmetry},\ }\href {https://doi.org/10.1103/PhysRevA.104.023318} {\bibfield  {journal} {\bibinfo  {journal} {Phys. Rev. A}\ }\textbf {\bibinfo {volume} {104}},\ \bibinfo {pages} {023318} (\bibinfo {year} {2021})}\BibitemShut {NoStop}%
\bibitem [{\citenamefont {Smits}\ \emph {et~al.}(2020)\citenamefont {Smits}, \citenamefont {Stoof},\ and\ \citenamefont {van~der Straten}}]{Smits_2020}%
  \BibitemOpen
  \bibfield  {author} {\bibinfo {author} {\bibfnamefont {J.}~\bibnamefont {Smits}}, \bibinfo {author} {\bibfnamefont {H.~T.~C.}\ \bibnamefont {Stoof}},\ and\ \bibinfo {author} {\bibfnamefont {P.}~\bibnamefont {van~der Straten}},\ }\bibfield  {title} {\bibinfo {title} {On the long-term stability of space-time crystals},\ }\href {https://doi.org/10.1088/1367-2630/abbae9} {\bibfield  {journal} {\bibinfo  {journal} {New Journal of Physics}\ }\textbf {\bibinfo {volume} {22}},\ \bibinfo {pages} {105001} (\bibinfo {year} {2020})}\BibitemShut {NoStop}%
\bibitem [{\citenamefont {Fu}\ \emph {et~al.}(2018)\citenamefont {Fu}, \citenamefont {Feng}, \citenamefont {Anderson}, \citenamefont {Clark}, \citenamefont {Hu}, \citenamefont {Andrade}, \citenamefont {Chin},\ and\ \citenamefont {Levin}}]{fu.2018.density}%
  \BibitemOpen
  \bibfield  {author} {\bibinfo {author} {\bibfnamefont {H.}~\bibnamefont {Fu}}, \bibinfo {author} {\bibfnamefont {L.}~\bibnamefont {Feng}}, \bibinfo {author} {\bibfnamefont {B.~M.}\ \bibnamefont {Anderson}}, \bibinfo {author} {\bibfnamefont {L.~W.}\ \bibnamefont {Clark}}, \bibinfo {author} {\bibfnamefont {J.}~\bibnamefont {Hu}}, \bibinfo {author} {\bibfnamefont {J.~W.}\ \bibnamefont {Andrade}}, \bibinfo {author} {\bibfnamefont {C.}~\bibnamefont {Chin}},\ and\ \bibinfo {author} {\bibfnamefont {K.}~\bibnamefont {Levin}},\ }\bibfield  {title} {\bibinfo {title} {Density waves and jet emission asymmetry in {Bose} fireworks},\ }\href {https://doi.org/10.1103/PhysRevLett.121.243001} {\bibfield  {journal} {\bibinfo  {journal} {Phys. Rev. Lett.}\ }\textbf {\bibinfo {volume} {121}},\ \bibinfo {pages} {243001} (\bibinfo {year} {2018})}\BibitemShut {NoStop}%
\bibitem [{\citenamefont {Kwon}\ \emph {et~al.}(2021)\citenamefont {Kwon}, \citenamefont {Mukherjee}, \citenamefont {Huh}, \citenamefont {Kim}, \citenamefont {Mistakidis}, \citenamefont {Maity}, \citenamefont {Kevrekidis}, \citenamefont {Majumder}, \citenamefont {Schmelcher},\ and\ \citenamefont {Choi}}]{kwon.2021.spontaneous}%
  \BibitemOpen
  \bibfield  {author} {\bibinfo {author} {\bibfnamefont {K.}~\bibnamefont {Kwon}}, \bibinfo {author} {\bibfnamefont {K.}~\bibnamefont {Mukherjee}}, \bibinfo {author} {\bibfnamefont {S.~J.}\ \bibnamefont {Huh}}, \bibinfo {author} {\bibfnamefont {K.}~\bibnamefont {Kim}}, \bibinfo {author} {\bibfnamefont {S.~I.}\ \bibnamefont {Mistakidis}}, \bibinfo {author} {\bibfnamefont {D.~K.}\ \bibnamefont {Maity}}, \bibinfo {author} {\bibfnamefont {P.~G.}\ \bibnamefont {Kevrekidis}}, \bibinfo {author} {\bibfnamefont {S.}~\bibnamefont {Majumder}}, \bibinfo {author} {\bibfnamefont {P.}~\bibnamefont {Schmelcher}},\ and\ \bibinfo {author} {\bibfnamefont {J.-y.}\ \bibnamefont {Choi}},\ }\bibfield  {title} {\bibinfo {title} {Spontaneous formation of star-shaped surface patterns in a driven {Bose}-{Einstein} condensate},\ }\href {https://doi.org/10.1103/PhysRevLett.127.113001} {\bibfield  {journal} {\bibinfo  {journal} {Phys. Rev. Lett.}\ }\textbf {\bibinfo {volume} {127}},\ \bibinfo {pages} {113001} (\bibinfo {year}
  {2021})}\BibitemShut {NoStop}%
\bibitem [{\citenamefont {Zhang}\ \emph {et~al.}(2020)\citenamefont {Zhang}, \citenamefont {Yao}, \citenamefont {Feng}, \citenamefont {Hu},\ and\ \citenamefont {Chin}}]{zhang.2020.pattern}%
  \BibitemOpen
  \bibfield  {author} {\bibinfo {author} {\bibfnamefont {Z.}~\bibnamefont {Zhang}}, \bibinfo {author} {\bibfnamefont {K.-X.}\ \bibnamefont {Yao}}, \bibinfo {author} {\bibfnamefont {L.}~\bibnamefont {Feng}}, \bibinfo {author} {\bibfnamefont {J.}~\bibnamefont {Hu}},\ and\ \bibinfo {author} {\bibfnamefont {C.}~\bibnamefont {Chin}},\ }\bibfield  {title} {\bibinfo {title} {Pattern formation in a driven {Bose}–{Einstein} condensate},\ }\href {https://doi.org/10.1038/s41567-020-0839-3} {\bibfield  {journal} {\bibinfo  {journal} {Nature Physics}\ }\textbf {\bibinfo {volume} {16}},\ \bibinfo {pages} {652} (\bibinfo {year} {2020})}\BibitemShut {NoStop}%
\bibitem [{\citenamefont {Liebster}\ \emph {et~al.}(2025)\citenamefont {Liebster}, \citenamefont {Sparn}, \citenamefont {Kath}, \citenamefont {Duchene}, \citenamefont {Fujii}, \citenamefont {G\"orlitz}, \citenamefont {Enss}, \citenamefont {Strobel},\ and\ \citenamefont {Oberthaler}}]{liebster.2025.prx}%
  \BibitemOpen
  \bibfield  {author} {\bibinfo {author} {\bibfnamefont {N.}~\bibnamefont {Liebster}}, \bibinfo {author} {\bibfnamefont {M.}~\bibnamefont {Sparn}}, \bibinfo {author} {\bibfnamefont {E.}~\bibnamefont {Kath}}, \bibinfo {author} {\bibfnamefont {J.}~\bibnamefont {Duchene}}, \bibinfo {author} {\bibfnamefont {K.}~\bibnamefont {Fujii}}, \bibinfo {author} {\bibfnamefont {S.~L.}\ \bibnamefont {G\"orlitz}}, \bibinfo {author} {\bibfnamefont {T.}~\bibnamefont {Enss}}, \bibinfo {author} {\bibfnamefont {H.}~\bibnamefont {Strobel}},\ and\ \bibinfo {author} {\bibfnamefont {M.~K.}\ \bibnamefont {Oberthaler}},\ }\bibfield  {title} {\bibinfo {title} {Observation of pattern stabilization in a driven superfluid},\ }\href {https://doi.org/10.1103/PhysRevX.15.011026} {\bibfield  {journal} {\bibinfo  {journal} {Phys. Rev. X}\ }\textbf {\bibinfo {volume} {15}},\ \bibinfo {pages} {011026} (\bibinfo {year} {2025})}\BibitemShut {NoStop}%
\bibitem [{\citenamefont {Gondret}\ \emph {et~al.}(2025{\natexlab{a}})\citenamefont {Gondret}, \citenamefont {Lamirault}, \citenamefont {Dias}, \citenamefont {Camier}, \citenamefont {Micheli}, \citenamefont {Leprince}, \citenamefont {Marolleau}, \citenamefont {Rullier}, \citenamefont {Robertson}, \citenamefont {Boiron},\ and\ \citenamefont {Westbrook}}]{gondret.2025.observation}%
  \BibitemOpen
  \bibfield  {author} {\bibinfo {author} {\bibfnamefont {V.}~\bibnamefont {Gondret}}, \bibinfo {author} {\bibfnamefont {C.}~\bibnamefont {Lamirault}}, \bibinfo {author} {\bibfnamefont {R.}~\bibnamefont {Dias}}, \bibinfo {author} {\bibfnamefont {L.}~\bibnamefont {Camier}}, \bibinfo {author} {\bibfnamefont {A.}~\bibnamefont {Micheli}}, \bibinfo {author} {\bibfnamefont {C.}~\bibnamefont {Leprince}}, \bibinfo {author} {\bibfnamefont {Q.}~\bibnamefont {Marolleau}}, \bibinfo {author} {\bibfnamefont {J.-R.}\ \bibnamefont {Rullier}}, \bibinfo {author} {\bibfnamefont {S.}~\bibnamefont {Robertson}}, \bibinfo {author} {\bibfnamefont {D.}~\bibnamefont {Boiron}},\ and\ \bibinfo {author} {\bibfnamefont {C.~I.}\ \bibnamefont {Westbrook}},\ }\bibfield  {title} {\bibinfo {title} {Observation of entanglement in a cold atom analog of cosmological preheating},\ }\href {https://doi.org/10.1103/h7ws-g9z2} {\bibfield  {journal} {\bibinfo  {journal} {Phys. Rev. Lett.}\ }\textbf {\bibinfo {volume} {135}},\ \bibinfo {pages} {240603}
  (\bibinfo {year} {2025}{\natexlab{a}})}\BibitemShut {NoStop}%
\bibitem [{\citenamefont {Capuzzi}\ and\ \citenamefont {Vignolo}(2008)}]{capuzzi.2008.faraday}%
  \BibitemOpen
  \bibfield  {author} {\bibinfo {author} {\bibfnamefont {P.}~\bibnamefont {Capuzzi}}\ and\ \bibinfo {author} {\bibfnamefont {P.}~\bibnamefont {Vignolo}},\ }\bibfield  {title} {\bibinfo {title} {{Faraday} waves in elongated superfluid fermionic clouds},\ }\href {https://doi.org/10.1103/PhysRevA.78.043613} {\bibfield  {journal} {\bibinfo  {journal} {Phys. Rev. A}\ }\textbf {\bibinfo {volume} {78}},\ \bibinfo {pages} {043613} (\bibinfo {year} {2008})}\BibitemShut {NoStop}%
\bibitem [{\citenamefont {Hern{\'a}ndez-Rajkov}\ \emph {et~al.}(2021)\citenamefont {Hern{\'a}ndez-Rajkov}, \citenamefont {Padilla-Castillo}, \citenamefont {del R{\'\i}o-Lima}, \citenamefont {Guti{\'e}rrez-Vald{\'e}s}, \citenamefont {Poveda-Cuevas},\ and\ \citenamefont {Seman}}]{hernandez.2021.faraday}%
  \BibitemOpen
  \bibfield  {author} {\bibinfo {author} {\bibfnamefont {D.}~\bibnamefont {Hern{\'a}ndez-Rajkov}}, \bibinfo {author} {\bibfnamefont {J.~E.}\ \bibnamefont {Padilla-Castillo}}, \bibinfo {author} {\bibfnamefont {A.}~\bibnamefont {del R{\'\i}o-Lima}}, \bibinfo {author} {\bibfnamefont {A.}~\bibnamefont {Guti{\'e}rrez-Vald{\'e}s}}, \bibinfo {author} {\bibfnamefont {F.~J.}\ \bibnamefont {Poveda-Cuevas}},\ and\ \bibinfo {author} {\bibfnamefont {J.~A.}\ \bibnamefont {Seman}},\ }\bibfield  {title} {\bibinfo {title} {{Faraday} waves in strongly interacting superfluids},\ }\href {https://doi.org/10.1088/1367-2630/ac2d70} {\bibfield  {journal} {\bibinfo  {journal} {New Journal of Physics}\ }\textbf {\bibinfo {volume} {23}},\ \bibinfo {pages} {103038} (\bibinfo {year} {2021})}\BibitemShut {NoStop}%
\bibitem [{\citenamefont {Cominotti}\ \emph {et~al.}(2022)\citenamefont {Cominotti}, \citenamefont {Berti}, \citenamefont {Farolfi}, \citenamefont {Zenesini}, \citenamefont {Lamporesi}, \citenamefont {Carusotto}, \citenamefont {Recati},\ and\ \citenamefont {Ferrari}}]{cominotti.2022.observation}%
  \BibitemOpen
  \bibfield  {author} {\bibinfo {author} {\bibfnamefont {R.}~\bibnamefont {Cominotti}}, \bibinfo {author} {\bibfnamefont {A.}~\bibnamefont {Berti}}, \bibinfo {author} {\bibfnamefont {A.}~\bibnamefont {Farolfi}}, \bibinfo {author} {\bibfnamefont {A.}~\bibnamefont {Zenesini}}, \bibinfo {author} {\bibfnamefont {G.}~\bibnamefont {Lamporesi}}, \bibinfo {author} {\bibfnamefont {I.}~\bibnamefont {Carusotto}}, \bibinfo {author} {\bibfnamefont {A.}~\bibnamefont {Recati}},\ and\ \bibinfo {author} {\bibfnamefont {G.}~\bibnamefont {Ferrari}},\ }\bibfield  {title} {\bibinfo {title} {Observation of massless and massive collective excitations with {Faraday} patterns in a two-component superfluid},\ }\href {https://doi.org/10.1103/PhysRevLett.128.210401} {\bibfield  {journal} {\bibinfo  {journal} {Phys. Rev. Lett.}\ }\textbf {\bibinfo {volume} {128}},\ \bibinfo {pages} {210401} (\bibinfo {year} {2022})}\BibitemShut {NoStop}%
\bibitem [{\citenamefont {\L{}akomy}\ \emph {et~al.}(2012)\citenamefont {\L{}akomy}, \citenamefont {Nath},\ and\ \citenamefont {Santos}}]{lakomy.2012.faraday}%
  \BibitemOpen
  \bibfield  {author} {\bibinfo {author} {\bibfnamefont {K.}~\bibnamefont {\L{}akomy}}, \bibinfo {author} {\bibfnamefont {R.}~\bibnamefont {Nath}},\ and\ \bibinfo {author} {\bibfnamefont {L.}~\bibnamefont {Santos}},\ }\bibfield  {title} {\bibinfo {title} {{Faraday} patterns in coupled one-dimensional dipolar condensates},\ }\href {https://doi.org/10.1103/PhysRevA.86.023620} {\bibfield  {journal} {\bibinfo  {journal} {Phys. Rev. A}\ }\textbf {\bibinfo {volume} {86}},\ \bibinfo {pages} {023620} (\bibinfo {year} {2012})}\BibitemShut {NoStop}%
\bibitem [{\citenamefont {Capuzzi}\ \emph {et~al.}(2011)\citenamefont {Capuzzi}, \citenamefont {Gattobigio},\ and\ \citenamefont {Vignolo}}]{capuzzi.2011.suppression}%
  \BibitemOpen
  \bibfield  {author} {\bibinfo {author} {\bibfnamefont {P.}~\bibnamefont {Capuzzi}}, \bibinfo {author} {\bibfnamefont {M.}~\bibnamefont {Gattobigio}},\ and\ \bibinfo {author} {\bibfnamefont {P.}~\bibnamefont {Vignolo}},\ }\bibfield  {title} {\bibinfo {title} {Suppression of {Faraday} waves in a {Bose}-{Einstein} condensate in the presence of an optical lattice},\ }\href {https://doi.org/10.1103/PhysRevA.83.013603} {\bibfield  {journal} {\bibinfo  {journal} {Phys. Rev. A}\ }\textbf {\bibinfo {volume} {83}},\ \bibinfo {pages} {013603} (\bibinfo {year} {2011})}\BibitemShut {NoStop}%
\bibitem [{\citenamefont {Dupont}\ \emph {et~al.}(2023)\citenamefont {Dupont}, \citenamefont {Gabardos}, \citenamefont {Arrouas}, \citenamefont {Chatelain}, \citenamefont {Arnal}, \citenamefont {Billy}, \citenamefont {Schlagheck}, \citenamefont {Peaudecerf},\ and\ \citenamefont {Guéry-Odelin}}]{dupont.2023.emergence}%
  \BibitemOpen
  \bibfield  {author} {\bibinfo {author} {\bibfnamefont {N.}~\bibnamefont {Dupont}}, \bibinfo {author} {\bibfnamefont {L.}~\bibnamefont {Gabardos}}, \bibinfo {author} {\bibfnamefont {F.}~\bibnamefont {Arrouas}}, \bibinfo {author} {\bibfnamefont {G.}~\bibnamefont {Chatelain}}, \bibinfo {author} {\bibfnamefont {M.}~\bibnamefont {Arnal}}, \bibinfo {author} {\bibfnamefont {J.}~\bibnamefont {Billy}}, \bibinfo {author} {\bibfnamefont {P.}~\bibnamefont {Schlagheck}}, \bibinfo {author} {\bibfnamefont {B.}~\bibnamefont {Peaudecerf}},\ and\ \bibinfo {author} {\bibfnamefont {D.}~\bibnamefont {Guéry-Odelin}},\ }\bibfield  {title} {\bibinfo {title} {Emergence of tunable periodic density correlations in a floquet–bloch system},\ }\href {https://doi.org/10.1073/pnas.2300980120} {\bibfield  {journal} {\bibinfo  {journal} {Proceedings of the National Academy of Sciences}\ }\textbf {\bibinfo {volume} {120}},\ \bibinfo {pages} {e2300980120} (\bibinfo {year} {2023})}\BibitemShut {NoStop}%
\bibitem [{\citenamefont {Unruh}(1981)}]{Unruh-1981}%
  \BibitemOpen
  \bibfield  {author} {\bibinfo {author} {\bibfnamefont {W.~G.}\ \bibnamefont {Unruh}},\ }\bibfield  {title} {\bibinfo {title} {Experimental black-hole evaporation?},\ }\href {https://doi.org/10.1103/PhysRevLett.46.1351} {\bibfield  {journal} {\bibinfo  {journal} {Phys. Rev. Lett.}\ }\textbf {\bibinfo {volume} {46}},\ \bibinfo {pages} {1351} (\bibinfo {year} {1981})}\BibitemShut {NoStop}%
\bibitem [{\citenamefont {Barceló}\ \emph {et~al.}(2011)\citenamefont {Barceló}, \citenamefont {Liberati},\ and\ \citenamefont {Visser}}]{Barcelo2011}%
  \BibitemOpen
  \bibfield  {author} {\bibinfo {author} {\bibfnamefont {C.}~\bibnamefont {Barceló}}, \bibinfo {author} {\bibfnamefont {S.}~\bibnamefont {Liberati}},\ and\ \bibinfo {author} {\bibfnamefont {M.}~\bibnamefont {Visser}},\ }\bibfield  {title} {\bibinfo {title} {Analogue gravity},\ }\href {https://doi.org/10.12942/lrr-2011-3} {\bibfield  {journal} {\bibinfo  {journal} {Living Reviews in Relativity}\ }\textbf {\bibinfo {volume} {14}},\ \bibinfo {pages} {3} (\bibinfo {year} {2011})}\BibitemShut {NoStop}%
\bibitem [{\citenamefont {Schützhold}(2025)}]{Schutzhold.2025.ultracoldatoms}%
  \BibitemOpen
  \bibfield  {author} {\bibinfo {author} {\bibfnamefont {R.}~\bibnamefont {Schützhold}},\ }\bibfield  {title} {\bibinfo {title} {Ultra-cold atoms as quantum simulators for relativistic phenomena},\ }\href {https://doi.org/https://doi.org/10.1016/j.ppnp.2025.104198} {\bibfield  {journal} {\bibinfo  {journal} {Progress in Particle and Nuclear Physics}\ }\textbf {\bibinfo {volume} {145}},\ \bibinfo {pages} {104198} (\bibinfo {year} {2025})}\BibitemShut {NoStop}%
\bibitem [{\citenamefont {Garay}\ \emph {et~al.}(2000)\citenamefont {Garay}, \citenamefont {Anglin}, \citenamefont {Cirac},\ and\ \citenamefont {Zoller}}]{garay.2000.sonic}%
  \BibitemOpen
  \bibfield  {author} {\bibinfo {author} {\bibfnamefont {L.~J.}\ \bibnamefont {Garay}}, \bibinfo {author} {\bibfnamefont {J.~R.}\ \bibnamefont {Anglin}}, \bibinfo {author} {\bibfnamefont {J.~I.}\ \bibnamefont {Cirac}},\ and\ \bibinfo {author} {\bibfnamefont {P.}~\bibnamefont {Zoller}},\ }\bibfield  {title} {\bibinfo {title} {Sonic analog of gravitational black holes in {Bose}-{Einstein} condensates},\ }\href {https://doi.org/10.1103/PhysRevLett.85.4643} {\bibfield  {journal} {\bibinfo  {journal} {Phys. Rev. Lett.}\ }\textbf {\bibinfo {volume} {85}},\ \bibinfo {pages} {4643} (\bibinfo {year} {2000})}\BibitemShut {NoStop}%
\bibitem [{\citenamefont {Fedichev}\ and\ \citenamefont {Fischer}(2004)}]{fedichev.2004.cosmological}%
  \BibitemOpen
  \bibfield  {author} {\bibinfo {author} {\bibfnamefont {P.~O.}\ \bibnamefont {Fedichev}}\ and\ \bibinfo {author} {\bibfnamefont {U.~R.}\ \bibnamefont {Fischer}},\ }\bibfield  {title} {\bibinfo {title} {``cosmological'' quasiparticle production in harmonically trapped superfluid gases},\ }\href {https://doi.org/10.1103/PhysRevA.69.033602} {\bibfield  {journal} {\bibinfo  {journal} {Phys. Rev. A}\ }\textbf {\bibinfo {volume} {69}},\ \bibinfo {pages} {033602} (\bibinfo {year} {2004})}\BibitemShut {NoStop}%
\bibitem [{\citenamefont {Fischer}\ and\ \citenamefont {Sch\"utzhold}(2004)}]{fischer.2004.quantum}%
  \BibitemOpen
  \bibfield  {author} {\bibinfo {author} {\bibfnamefont {U.~R.}\ \bibnamefont {Fischer}}\ and\ \bibinfo {author} {\bibfnamefont {R.}~\bibnamefont {Sch\"utzhold}},\ }\bibfield  {title} {\bibinfo {title} {Quantum simulation of cosmic inflation in two-component {Bose}-{Einstein} condensates},\ }\href {https://doi.org/10.1103/PhysRevA.70.063615} {\bibfield  {journal} {\bibinfo  {journal} {Phys. Rev. A}\ }\textbf {\bibinfo {volume} {70}},\ \bibinfo {pages} {063615} (\bibinfo {year} {2004})}\BibitemShut {NoStop}%
\bibitem [{\citenamefont {Uhlmann}\ \emph {et~al.}(2005)\citenamefont {Uhlmann}, \citenamefont {Xu},\ and\ \citenamefont {Schützhold}}]{Uhlmann_2005}%
  \BibitemOpen
  \bibfield  {author} {\bibinfo {author} {\bibfnamefont {M.}~\bibnamefont {Uhlmann}}, \bibinfo {author} {\bibfnamefont {Y.}~\bibnamefont {Xu}},\ and\ \bibinfo {author} {\bibfnamefont {R.}~\bibnamefont {Schützhold}},\ }\bibfield  {title} {\bibinfo {title} {Aspects of cosmic inflation in expanding {Bose}–{Einstein} condensates},\ }\href {https://doi.org/10.1088/1367-2630/7/1/248} {\bibfield  {journal} {\bibinfo  {journal} {New Journal of Physics}\ }\textbf {\bibinfo {volume} {7}},\ \bibinfo {pages} {248} (\bibinfo {year} {2005})}\BibitemShut {NoStop}%
\bibitem [{\citenamefont {Jain}\ \emph {et~al.}(2007)\citenamefont {Jain}, \citenamefont {Weinfurtner}, \citenamefont {Visser},\ and\ \citenamefont {Gardiner}}]{piyush.2007.analog}%
  \BibitemOpen
  \bibfield  {author} {\bibinfo {author} {\bibfnamefont {P.}~\bibnamefont {Jain}}, \bibinfo {author} {\bibfnamefont {S.}~\bibnamefont {Weinfurtner}}, \bibinfo {author} {\bibfnamefont {M.}~\bibnamefont {Visser}},\ and\ \bibinfo {author} {\bibfnamefont {C.~W.}\ \bibnamefont {Gardiner}},\ }\bibfield  {title} {\bibinfo {title} {Analog model of a {Friedmann-Robertson-Walker} universe in {Bose}-{Einstein} condensates: Application of the classical field method},\ }\href {https://doi.org/10.1103/PhysRevA.76.033616} {\bibfield  {journal} {\bibinfo  {journal} {Phys. Rev. A}\ }\textbf {\bibinfo {volume} {76}},\ \bibinfo {pages} {033616} (\bibinfo {year} {2007})}\BibitemShut {NoStop}%
\bibitem [{\citenamefont {Carusotto}\ \emph {et~al.}(2010)\citenamefont {Carusotto}, \citenamefont {Balbinot}, \citenamefont {Fabbri},\ and\ \citenamefont {Recati}}]{carusotto.2010.density}%
  \BibitemOpen
  \bibfield  {author} {\bibinfo {author} {\bibfnamefont {I.}~\bibnamefont {Carusotto}}, \bibinfo {author} {\bibfnamefont {R.}~\bibnamefont {Balbinot}}, \bibinfo {author} {\bibfnamefont {A.}~\bibnamefont {Fabbri}},\ and\ \bibinfo {author} {\bibfnamefont {A.}~\bibnamefont {Recati}},\ }\bibfield  {title} {\bibinfo {title} {Density correlations and analog dynamical {Casimir} emission of {Bogoliubov} phonons in modulated atomic {Bose}-{Einstein} condensates},\ }\href {https://doi.org/10.1140/epjd/e2009-00314-3} {\bibfield  {journal} {\bibinfo  {journal} {The European Physical Journal D}\ }\textbf {\bibinfo {volume} {56}},\ \bibinfo {pages} {391} (\bibinfo {year} {2010})}\BibitemShut {NoStop}%
\bibitem [{\citenamefont {Tian}\ \emph {et~al.}(2018)\citenamefont {Tian}, \citenamefont {Ch\"a},\ and\ \citenamefont {Fischer}}]{tian.2018.roton}%
  \BibitemOpen
  \bibfield  {author} {\bibinfo {author} {\bibfnamefont {Z.}~\bibnamefont {Tian}}, \bibinfo {author} {\bibfnamefont {S.-Y.}\ \bibnamefont {Ch\"a}},\ and\ \bibinfo {author} {\bibfnamefont {U.~R.}\ \bibnamefont {Fischer}},\ }\bibfield  {title} {\bibinfo {title} {Roton entanglement in quenched dipolar {Bose}-{Einstein} condensates},\ }\href {https://doi.org/10.1103/PhysRevA.97.063611} {\bibfield  {journal} {\bibinfo  {journal} {Phys. Rev. A}\ }\textbf {\bibinfo {volume} {97}},\ \bibinfo {pages} {063611} (\bibinfo {year} {2018})}\BibitemShut {NoStop}%
\bibitem [{\citenamefont {Macher}\ and\ \citenamefont {Parentani}(2009)}]{macher.2009.black-hole}%
  \BibitemOpen
  \bibfield  {author} {\bibinfo {author} {\bibfnamefont {J.}~\bibnamefont {Macher}}\ and\ \bibinfo {author} {\bibfnamefont {R.}~\bibnamefont {Parentani}},\ }\bibfield  {title} {\bibinfo {title} {Black-hole radiation in {Bose}-{Einstein} condensates},\ }\href {https://doi.org/10.1103/PhysRevA.80.043601} {\bibfield  {journal} {\bibinfo  {journal} {Phys. Rev. A}\ }\textbf {\bibinfo {volume} {80}},\ \bibinfo {pages} {043601} (\bibinfo {year} {2009})}\BibitemShut {NoStop}%
\bibitem [{\citenamefont {Busch}\ and\ \citenamefont {Parentani}(2013)}]{busch.2013.dce.dissipative}%
  \BibitemOpen
  \bibfield  {author} {\bibinfo {author} {\bibfnamefont {X.}~\bibnamefont {Busch}}\ and\ \bibinfo {author} {\bibfnamefont {R.}~\bibnamefont {Parentani}},\ }\bibfield  {title} {\bibinfo {title} {Dynamical {Casimir} effect in dissipative media: When is the final state nonseparable?},\ }\href {https://doi.org/10.1103/PhysRevD.88.045023} {\bibfield  {journal} {\bibinfo  {journal} {Phys. Rev. D}\ }\textbf {\bibinfo {volume} {88}},\ \bibinfo {pages} {045023} (\bibinfo {year} {2013})}\BibitemShut {NoStop}%
\bibitem [{\citenamefont {Busch}\ \emph {et~al.}(2014)\citenamefont {Busch}, \citenamefont {Parentani},\ and\ \citenamefont {Robertson}}]{busch.2014.dce.quantum}%
  \BibitemOpen
  \bibfield  {author} {\bibinfo {author} {\bibfnamefont {X.}~\bibnamefont {Busch}}, \bibinfo {author} {\bibfnamefont {R.}~\bibnamefont {Parentani}},\ and\ \bibinfo {author} {\bibfnamefont {S.}~\bibnamefont {Robertson}},\ }\bibfield  {title} {\bibinfo {title} {Quantum entanglement due to a modulated dynamical {Casimir} effect},\ }\href {https://doi.org/10.1103/PhysRevA.89.063606} {\bibfield  {journal} {\bibinfo  {journal} {Phys. Rev. A}\ }\textbf {\bibinfo {volume} {89}},\ \bibinfo {pages} {063606} (\bibinfo {year} {2014})}\BibitemShut {NoStop}%
\bibitem [{\citenamefont {Robertson}\ \emph {et~al.}(2017{\natexlab{a}})\citenamefont {Robertson}, \citenamefont {Michel},\ and\ \citenamefont {Parentani}}]{robertson.2017.controlling}%
  \BibitemOpen
  \bibfield  {author} {\bibinfo {author} {\bibfnamefont {S.}~\bibnamefont {Robertson}}, \bibinfo {author} {\bibfnamefont {F.}~\bibnamefont {Michel}},\ and\ \bibinfo {author} {\bibfnamefont {R.}~\bibnamefont {Parentani}},\ }\bibfield  {title} {\bibinfo {title} {Controlling and observing nonseparability of phonons created in time-dependent {1D} atomic {Bose} condensates},\ }\href {https://doi.org/10.1103/PhysRevD.95.065020} {\bibfield  {journal} {\bibinfo  {journal} {Phys. Rev. D}\ }\textbf {\bibinfo {volume} {95}},\ \bibinfo {pages} {065020} (\bibinfo {year} {2017}{\natexlab{a}})}\BibitemShut {NoStop}%
\bibitem [{\citenamefont {Robertson}\ \emph {et~al.}(2017{\natexlab{b}})\citenamefont {Robertson}, \citenamefont {Michel},\ and\ \citenamefont {Parentani}}]{robertson.2017.assessing}%
  \BibitemOpen
  \bibfield  {author} {\bibinfo {author} {\bibfnamefont {S.}~\bibnamefont {Robertson}}, \bibinfo {author} {\bibfnamefont {F.}~\bibnamefont {Michel}},\ and\ \bibinfo {author} {\bibfnamefont {R.}~\bibnamefont {Parentani}},\ }\bibfield  {title} {\bibinfo {title} {Assessing degrees of entanglement of phonon states in atomic {Bose} gases through the measurement of commuting observables},\ }\href {https://doi.org/10.1103/PhysRevD.96.045012} {\bibfield  {journal} {\bibinfo  {journal} {Phys. Rev. D}\ }\textbf {\bibinfo {volume} {96}},\ \bibinfo {pages} {045012} (\bibinfo {year} {2017}{\natexlab{b}})}\BibitemShut {NoStop}%
\bibitem [{\citenamefont {Wilson}\ \emph {et~al.}(2011)\citenamefont {Wilson}, \citenamefont {Johansson}, \citenamefont {Pourkabirian}, \citenamefont {Simoen}, \citenamefont {Johansson}, \citenamefont {Duty}, \citenamefont {Nori},\ and\ \citenamefont {Delsing}}]{wilson.2011.observation}%
  \BibitemOpen
  \bibfield  {author} {\bibinfo {author} {\bibfnamefont {C.~M.}\ \bibnamefont {Wilson}}, \bibinfo {author} {\bibfnamefont {G.}~\bibnamefont {Johansson}}, \bibinfo {author} {\bibfnamefont {A.}~\bibnamefont {Pourkabirian}}, \bibinfo {author} {\bibfnamefont {M.}~\bibnamefont {Simoen}}, \bibinfo {author} {\bibfnamefont {J.~R.}\ \bibnamefont {Johansson}}, \bibinfo {author} {\bibfnamefont {T.}~\bibnamefont {Duty}}, \bibinfo {author} {\bibfnamefont {F.}~\bibnamefont {Nori}},\ and\ \bibinfo {author} {\bibfnamefont {P.}~\bibnamefont {Delsing}},\ }\bibfield  {title} {\bibinfo {title} {Observation of the dynamical {Casimir} effect in a superconducting circuit},\ }\href {https://doi.org/10.1038/nature10561} {\bibfield  {journal} {\bibinfo  {journal} {Nature}\ }\textbf {\bibinfo {volume} {479}},\ \bibinfo {pages} {376} (\bibinfo {year} {2011})}\BibitemShut {NoStop}%
\bibitem [{\citenamefont {Jaskula}\ \emph {et~al.}(2012)\citenamefont {Jaskula}, \citenamefont {Partridge}, \citenamefont {Bonneau}, \citenamefont {Lopes}, \citenamefont {Ruaudel}, \citenamefont {Boiron},\ and\ \citenamefont {Westbrook}}]{jaskula.2012.acoustic}%
  \BibitemOpen
  \bibfield  {author} {\bibinfo {author} {\bibfnamefont {J.-C.}\ \bibnamefont {Jaskula}}, \bibinfo {author} {\bibfnamefont {G.~B.}\ \bibnamefont {Partridge}}, \bibinfo {author} {\bibfnamefont {M.}~\bibnamefont {Bonneau}}, \bibinfo {author} {\bibfnamefont {R.}~\bibnamefont {Lopes}}, \bibinfo {author} {\bibfnamefont {J.}~\bibnamefont {Ruaudel}}, \bibinfo {author} {\bibfnamefont {D.}~\bibnamefont {Boiron}},\ and\ \bibinfo {author} {\bibfnamefont {C.~I.}\ \bibnamefont {Westbrook}},\ }\bibfield  {title} {\bibinfo {title} {Acoustic analog to the dynamical {Casimir} effect in a {Bose}-{Einstein} condensate},\ }\href {https://doi.org/10.1103/PhysRevLett.109.220401} {\bibfield  {journal} {\bibinfo  {journal} {Phys. Rev. Lett.}\ }\textbf {\bibinfo {volume} {109}},\ \bibinfo {pages} {220401} (\bibinfo {year} {2012})}\BibitemShut {NoStop}%
\bibitem [{\citenamefont {L{\"a}hteenm{\"a}ki}\ \emph {et~al.}(2013)\citenamefont {L{\"a}hteenm{\"a}ki}, \citenamefont {Paraoanu}, \citenamefont {Hassel},\ and\ \citenamefont {Hakonen}}]{Lahteenmaki:2011cwo}%
  \BibitemOpen
  \bibfield  {author} {\bibinfo {author} {\bibfnamefont {P.}~\bibnamefont {L{\"a}hteenm{\"a}ki}}, \bibinfo {author} {\bibfnamefont {G.~S.}\ \bibnamefont {Paraoanu}}, \bibinfo {author} {\bibfnamefont {J.}~\bibnamefont {Hassel}},\ and\ \bibinfo {author} {\bibfnamefont {P.~J.}\ \bibnamefont {Hakonen}},\ }\bibfield  {title} {\bibinfo {title} {Dynamical {Casimir} effect in a {Josephson} metamaterial},\ }\href {https://doi.org/10.1073/pnas.1212705110} {\bibfield  {journal} {\bibinfo  {journal} {Proceedings of the National Academy of Sciences}\ }\textbf {\bibinfo {volume} {110}},\ \bibinfo {pages} {4234} (\bibinfo {year} {2013})}\BibitemShut {NoStop}%
\bibitem [{\citenamefont {Vezzoli}\ \emph {et~al.}(2019)\citenamefont {Vezzoli}, \citenamefont {Mussot}, \citenamefont {Westerberg}, \citenamefont {Kudlinski}, \citenamefont {Dinparasti~Saleh}, \citenamefont {Prain}, \citenamefont {Biancalana}, \citenamefont {Lantz},\ and\ \citenamefont {Faccio}}]{vezzoli.2019.optical}%
  \BibitemOpen
  \bibfield  {author} {\bibinfo {author} {\bibfnamefont {S.}~\bibnamefont {Vezzoli}}, \bibinfo {author} {\bibfnamefont {A.}~\bibnamefont {Mussot}}, \bibinfo {author} {\bibfnamefont {N.}~\bibnamefont {Westerberg}}, \bibinfo {author} {\bibfnamefont {A.}~\bibnamefont {Kudlinski}}, \bibinfo {author} {\bibfnamefont {H.}~\bibnamefont {Dinparasti~Saleh}}, \bibinfo {author} {\bibfnamefont {A.}~\bibnamefont {Prain}}, \bibinfo {author} {\bibfnamefont {F.}~\bibnamefont {Biancalana}}, \bibinfo {author} {\bibfnamefont {E.}~\bibnamefont {Lantz}},\ and\ \bibinfo {author} {\bibfnamefont {D.}~\bibnamefont {Faccio}},\ }\bibfield  {title} {\bibinfo {title} {Optical analogue of the dynamical {Casimir} effect in a dispersion-oscillating fibre},\ }\href {https://doi.org/10.1038/s42005-019-0183-z} {\bibfield  {journal} {\bibinfo  {journal} {Communications Physics}\ }\textbf {\bibinfo {volume} {2}},\ \bibinfo {pages} {84} (\bibinfo {year} {2019})}\BibitemShut {NoStop}%
\bibitem [{\citenamefont {Hu}\ \emph {et~al.}(2019)\citenamefont {Hu}, \citenamefont {Feng}, \citenamefont {Zhang},\ and\ \citenamefont {Chin}}]{hu.2019.quantum}%
  \BibitemOpen
  \bibfield  {author} {\bibinfo {author} {\bibfnamefont {J.}~\bibnamefont {Hu}}, \bibinfo {author} {\bibfnamefont {L.}~\bibnamefont {Feng}}, \bibinfo {author} {\bibfnamefont {Z.}~\bibnamefont {Zhang}},\ and\ \bibinfo {author} {\bibfnamefont {C.}~\bibnamefont {Chin}},\ }\bibfield  {title} {\bibinfo {title} {Quantum simulation of {Unruh} radiation},\ }\href {https://doi.org/10.1038/s41567-019-0537-1} {\bibfield  {journal} {\bibinfo  {journal} {Nature Physics}\ }\textbf {\bibinfo {volume} {15}},\ \bibinfo {pages} {785} (\bibinfo {year} {2019})}\BibitemShut {NoStop}%
\bibitem [{\citenamefont {Chen}\ \emph {et~al.}(2021)\citenamefont {Chen}, \citenamefont {Khlebnikov},\ and\ \citenamefont {Hung}}]{chen.2021.observation}%
  \BibitemOpen
  \bibfield  {author} {\bibinfo {author} {\bibfnamefont {C.-A.}\ \bibnamefont {Chen}}, \bibinfo {author} {\bibfnamefont {S.}~\bibnamefont {Khlebnikov}},\ and\ \bibinfo {author} {\bibfnamefont {C.-L.}\ \bibnamefont {Hung}},\ }\bibfield  {title} {\bibinfo {title} {Observation of quasiparticle pair production and quantum entanglement in atomic quantum gases quenched to an attractive interaction},\ }\href {https://doi.org/10.1103/PhysRevLett.127.060404} {\bibfield  {journal} {\bibinfo  {journal} {Phys. Rev. Lett.}\ }\textbf {\bibinfo {volume} {127}},\ \bibinfo {pages} {060404} (\bibinfo {year} {2021})}\BibitemShut {NoStop}%
\bibitem [{\citenamefont {Viermann}\ \emph {et~al.}(2022)\citenamefont {Viermann}, \citenamefont {Sparn}, \citenamefont {Liebster}, \citenamefont {Hans}, \citenamefont {Kath}, \citenamefont {Parra-L\'opez}, \citenamefont {Tolosa-Sime\'on}, \citenamefont {Sánchez-Kuntz}, \citenamefont {Haas}, \citenamefont {Strobel}, \citenamefont {Floerchinger},\ and\ \citenamefont {Oberthaler}}]{viermann_quantum_2022}%
  \BibitemOpen
  \bibfield  {author} {\bibinfo {author} {\bibfnamefont {C.}~\bibnamefont {Viermann}}, \bibinfo {author} {\bibfnamefont {M.}~\bibnamefont {Sparn}}, \bibinfo {author} {\bibfnamefont {N.}~\bibnamefont {Liebster}}, \bibinfo {author} {\bibfnamefont {M.}~\bibnamefont {Hans}}, \bibinfo {author} {\bibfnamefont {E.}~\bibnamefont {Kath}}, \bibinfo {author} {\bibfnamefont {A.}~\bibnamefont {Parra-L\'opez}}, \bibinfo {author} {\bibfnamefont {M.}~\bibnamefont {Tolosa-Sime\'on}}, \bibinfo {author} {\bibfnamefont {N.}~\bibnamefont {Sánchez-Kuntz}}, \bibinfo {author} {\bibfnamefont {T.}~\bibnamefont {Haas}}, \bibinfo {author} {\bibfnamefont {H.}~\bibnamefont {Strobel}}, \bibinfo {author} {\bibfnamefont {S.}~\bibnamefont {Floerchinger}},\ and\ \bibinfo {author} {\bibfnamefont {M.~K.}\ \bibnamefont {Oberthaler}},\ }\bibfield  {title} {\bibinfo {title} {Quantum field simulator for dynamics in curved spacetime},\ }\href {https://doi.org/10.1038/s41586-022-05313-9} {\bibfield  {journal} {\bibinfo  {journal} {Nature}\ }\textbf
  {\bibinfo {volume} {611}},\ \bibinfo {pages} {260} (\bibinfo {year} {2022})}\BibitemShut {NoStop}%
\bibitem [{\citenamefont {Steinhauer}\ \emph {et~al.}(2022)\citenamefont {Steinhauer}, \citenamefont {Abuzarli}, \citenamefont {Aladjidi}, \citenamefont {Bienaimé}, \citenamefont {Piekarski}, \citenamefont {Liu}, \citenamefont {Giacobino}, \citenamefont {Bramati},\ and\ \citenamefont {Glorieux}}]{steinhauer_analogue_2022}%
  \BibitemOpen
  \bibfield  {author} {\bibinfo {author} {\bibfnamefont {J.}~\bibnamefont {Steinhauer}}, \bibinfo {author} {\bibfnamefont {M.}~\bibnamefont {Abuzarli}}, \bibinfo {author} {\bibfnamefont {T.}~\bibnamefont {Aladjidi}}, \bibinfo {author} {\bibfnamefont {T.}~\bibnamefont {Bienaimé}}, \bibinfo {author} {\bibfnamefont {C.}~\bibnamefont {Piekarski}}, \bibinfo {author} {\bibfnamefont {W.}~\bibnamefont {Liu}}, \bibinfo {author} {\bibfnamefont {E.}~\bibnamefont {Giacobino}}, \bibinfo {author} {\bibfnamefont {A.}~\bibnamefont {Bramati}},\ and\ \bibinfo {author} {\bibfnamefont {Q.}~\bibnamefont {Glorieux}},\ }\bibfield  {title} {\bibinfo {title} {Analogue cosmological particle creation in an ultracold quantum fluid of light},\ }\href {https://doi.org/10.1038/s41467-022-30603-1} {\bibfield  {journal} {\bibinfo  {journal} {Nature Communications}\ }\textbf {\bibinfo {volume} {13}},\ \bibinfo {pages} {2890} (\bibinfo {year} {2022})}\BibitemShut {NoStop}%
\bibitem [{\citenamefont {Sparn}\ \emph {et~al.}(2024)\citenamefont {Sparn}, \citenamefont {Kath}, \citenamefont {Liebster}, \citenamefont {Duchene}, \citenamefont {Schmidt}, \citenamefont {Tolosa-Sime\'on}, \citenamefont {Parra-L\'opez}, \citenamefont {Floerchinger}, \citenamefont {Strobel},\ and\ \citenamefont {Oberthaler}}]{sparn.2024.experimental}%
  \BibitemOpen
  \bibfield  {author} {\bibinfo {author} {\bibfnamefont {M.}~\bibnamefont {Sparn}}, \bibinfo {author} {\bibfnamefont {E.}~\bibnamefont {Kath}}, \bibinfo {author} {\bibfnamefont {N.}~\bibnamefont {Liebster}}, \bibinfo {author} {\bibfnamefont {J.}~\bibnamefont {Duchene}}, \bibinfo {author} {\bibfnamefont {C.~F.}\ \bibnamefont {Schmidt}}, \bibinfo {author} {\bibfnamefont {M.}~\bibnamefont {Tolosa-Sime\'on}}, \bibinfo {author} {\bibfnamefont {A.}~\bibnamefont {Parra-L\'opez}}, \bibinfo {author} {\bibfnamefont {S.}~\bibnamefont {Floerchinger}}, \bibinfo {author} {\bibfnamefont {H.}~\bibnamefont {Strobel}},\ and\ \bibinfo {author} {\bibfnamefont {M.~K.}\ \bibnamefont {Oberthaler}},\ }\bibfield  {title} {\bibinfo {title} {Experimental particle production in time-dependent spacetimes: A one-dimensional scattering problem},\ }\href {https://doi.org/10.1103/PhysRevLett.133.260201} {\bibfield  {journal} {\bibinfo  {journal} {Phys. Rev. Lett.}\ }\textbf {\bibinfo {volume} {133}},\ \bibinfo {pages} {260201} (\bibinfo
  {year} {2024})}\BibitemShut {NoStop}%
\bibitem [{\citenamefont {Philbin}\ \emph {et~al.}(2008)\citenamefont {Philbin}, \citenamefont {Kuklewicz}, \citenamefont {Robertson}, \citenamefont {Hill}, \citenamefont {König},\ and\ \citenamefont {Leonhardt}}]{philbin.2008.fiber}%
  \BibitemOpen
  \bibfield  {author} {\bibinfo {author} {\bibfnamefont {T.~G.}\ \bibnamefont {Philbin}}, \bibinfo {author} {\bibfnamefont {C.}~\bibnamefont {Kuklewicz}}, \bibinfo {author} {\bibfnamefont {S.}~\bibnamefont {Robertson}}, \bibinfo {author} {\bibfnamefont {S.}~\bibnamefont {Hill}}, \bibinfo {author} {\bibfnamefont {F.}~\bibnamefont {König}},\ and\ \bibinfo {author} {\bibfnamefont {U.}~\bibnamefont {Leonhardt}},\ }\bibfield  {title} {\bibinfo {title} {Fiber-optical analog of the event horizon},\ }\href {https://doi.org/10.1126/science.1153625} {\bibfield  {journal} {\bibinfo  {journal} {Science}\ }\textbf {\bibinfo {volume} {319}},\ \bibinfo {pages} {1367} (\bibinfo {year} {2008})}\BibitemShut {NoStop}%
\bibitem [{\citenamefont {Weinfurtner}\ \emph {et~al.}(2011)\citenamefont {Weinfurtner}, \citenamefont {Tedford}, \citenamefont {Penrice}, \citenamefont {Unruh},\ and\ \citenamefont {Lawrence}}]{weinfurtner.2011.measurement}%
  \BibitemOpen
  \bibfield  {author} {\bibinfo {author} {\bibfnamefont {S.}~\bibnamefont {Weinfurtner}}, \bibinfo {author} {\bibfnamefont {E.~W.}\ \bibnamefont {Tedford}}, \bibinfo {author} {\bibfnamefont {M.~C.~J.}\ \bibnamefont {Penrice}}, \bibinfo {author} {\bibfnamefont {W.~G.}\ \bibnamefont {Unruh}},\ and\ \bibinfo {author} {\bibfnamefont {G.~A.}\ \bibnamefont {Lawrence}},\ }\bibfield  {title} {\bibinfo {title} {Measurement of stimulated {Hawking} emission in an analogue system},\ }\href {https://doi.org/10.1103/PhysRevLett.106.021302} {\bibfield  {journal} {\bibinfo  {journal} {Phys. Rev. Lett.}\ }\textbf {\bibinfo {volume} {106}},\ \bibinfo {pages} {021302} (\bibinfo {year} {2011})}\BibitemShut {NoStop}%
\bibitem [{\citenamefont {Steinhauer}(2016)}]{steinhauer.2016.observation}%
  \BibitemOpen
  \bibfield  {author} {\bibinfo {author} {\bibfnamefont {J.}~\bibnamefont {Steinhauer}},\ }\bibfield  {title} {\bibinfo {title} {Observation of quantum {Hawking} radiation and its entanglement in an analogue black hole},\ }\href {https://doi.org/10.1038/nphys3863} {\bibfield  {journal} {\bibinfo  {journal} {Nature Physics}\ }\textbf {\bibinfo {volume} {12}},\ \bibinfo {pages} {959} (\bibinfo {year} {2016})}\BibitemShut {NoStop}%
\bibitem [{\citenamefont {Euv\'e}\ \emph {et~al.}(2016)\citenamefont {Euv\'e}, \citenamefont {Michel}, \citenamefont {Parentani}, \citenamefont {Philbin},\ and\ \citenamefont {Rousseaux}}]{euve.2016.observation}%
  \BibitemOpen
  \bibfield  {author} {\bibinfo {author} {\bibfnamefont {L.-P.}\ \bibnamefont {Euv\'e}}, \bibinfo {author} {\bibfnamefont {F.}~\bibnamefont {Michel}}, \bibinfo {author} {\bibfnamefont {R.}~\bibnamefont {Parentani}}, \bibinfo {author} {\bibfnamefont {T.~G.}\ \bibnamefont {Philbin}},\ and\ \bibinfo {author} {\bibfnamefont {G.}~\bibnamefont {Rousseaux}},\ }\bibfield  {title} {\bibinfo {title} {Observation of noise correlated by the {Hawking} effect in a water tank},\ }\href {https://doi.org/10.1103/PhysRevLett.117.121301} {\bibfield  {journal} {\bibinfo  {journal} {Phys. Rev. Lett.}\ }\textbf {\bibinfo {volume} {117}},\ \bibinfo {pages} {121301} (\bibinfo {year} {2016})}\BibitemShut {NoStop}%
\bibitem [{\citenamefont {Muñoz De~Nova}\ \emph {et~al.}(2019)\citenamefont {Muñoz De~Nova}, \citenamefont {Golubkov}, \citenamefont {Kolobov},\ and\ \citenamefont {Steinhauer}}]{munoz_de_nova_observation_2019}%
  \BibitemOpen
  \bibfield  {author} {\bibinfo {author} {\bibfnamefont {J.~R.}\ \bibnamefont {Muñoz De~Nova}}, \bibinfo {author} {\bibfnamefont {K.}~\bibnamefont {Golubkov}}, \bibinfo {author} {\bibfnamefont {V.~I.}\ \bibnamefont {Kolobov}},\ and\ \bibinfo {author} {\bibfnamefont {J.}~\bibnamefont {Steinhauer}},\ }\bibfield  {title} {\bibinfo {title} {Observation of thermal {Hawking} radiation and its temperature in an analogue black hole},\ }\href {https://doi.org/10.1038/s41586-019-1241-0} {\bibfield  {journal} {\bibinfo  {journal} {Nature}\ }\textbf {\bibinfo {volume} {569}},\ \bibinfo {pages} {688} (\bibinfo {year} {2019})}\BibitemShut {NoStop}%
\bibitem [{\citenamefont {Drori}\ \emph {et~al.}(2019)\citenamefont {Drori}, \citenamefont {Rosenberg}, \citenamefont {Bermudez}, \citenamefont {Silberberg},\ and\ \citenamefont {Leonhardt}}]{observation.2019.drori}%
  \BibitemOpen
  \bibfield  {author} {\bibinfo {author} {\bibfnamefont {J.}~\bibnamefont {Drori}}, \bibinfo {author} {\bibfnamefont {Y.}~\bibnamefont {Rosenberg}}, \bibinfo {author} {\bibfnamefont {D.}~\bibnamefont {Bermudez}}, \bibinfo {author} {\bibfnamefont {Y.}~\bibnamefont {Silberberg}},\ and\ \bibinfo {author} {\bibfnamefont {U.}~\bibnamefont {Leonhardt}},\ }\bibfield  {title} {\bibinfo {title} {Observation of stimulated {Hawking} radiation in an optical analogue},\ }\href {https://doi.org/10.1103/PhysRevLett.122.010404} {\bibfield  {journal} {\bibinfo  {journal} {Phys. Rev. Lett.}\ }\textbf {\bibinfo {volume} {122}},\ \bibinfo {pages} {010404} (\bibinfo {year} {2019})}\BibitemShut {NoStop}%
\bibitem [{\citenamefont {Švančara}\ \emph {et~al.}(2024)\citenamefont {Švančara}, \citenamefont {Smaniotto}, \citenamefont {Solidoro}, \citenamefont {MacDonald}, \citenamefont {Patrick}, \citenamefont {Gregory}, \citenamefont {Barenghi},\ and\ \citenamefont {Weinfurtner}}]{svancara_rotating_2024}%
  \BibitemOpen
  \bibfield  {author} {\bibinfo {author} {\bibfnamefont {P.}~\bibnamefont {Švančara}}, \bibinfo {author} {\bibfnamefont {P.}~\bibnamefont {Smaniotto}}, \bibinfo {author} {\bibfnamefont {L.}~\bibnamefont {Solidoro}}, \bibinfo {author} {\bibfnamefont {J.~F.}\ \bibnamefont {MacDonald}}, \bibinfo {author} {\bibfnamefont {S.}~\bibnamefont {Patrick}}, \bibinfo {author} {\bibfnamefont {R.}~\bibnamefont {Gregory}}, \bibinfo {author} {\bibfnamefont {C.~F.}\ \bibnamefont {Barenghi}},\ and\ \bibinfo {author} {\bibfnamefont {S.}~\bibnamefont {Weinfurtner}},\ }\bibfield  {title} {\bibinfo {title} {Rotating curved spacetime signatures from a giant quantum vortex},\ }\href {https://doi.org/10.1038/s41586-024-07176-8} {\bibfield  {journal} {\bibinfo  {journal} {Nature}\ }\textbf {\bibinfo {volume} {628}},\ \bibinfo {pages} {66} (\bibinfo {year} {2024})}\BibitemShut {NoStop}%
\bibitem [{\citenamefont {Falque}\ \emph {et~al.}(2025)\citenamefont {Falque}, \citenamefont {Delhom}, \citenamefont {Glorieux}, \citenamefont {Giacobino}, \citenamefont {Bramati},\ and\ \citenamefont {Jacquet}}]{falque.2025.polariton}%
  \BibitemOpen
  \bibfield  {author} {\bibinfo {author} {\bibfnamefont {K.}~\bibnamefont {Falque}}, \bibinfo {author} {\bibfnamefont {A.}~\bibnamefont {Delhom}}, \bibinfo {author} {\bibfnamefont {Q.}~\bibnamefont {Glorieux}}, \bibinfo {author} {\bibfnamefont {E.}~\bibnamefont {Giacobino}}, \bibinfo {author} {\bibfnamefont {A.}~\bibnamefont {Bramati}},\ and\ \bibinfo {author} {\bibfnamefont {M.~J.}\ \bibnamefont {Jacquet}},\ }\bibfield  {title} {\bibinfo {title} {Polariton fluids as quantum field theory simulators on tailored curved spacetimes},\ }\href {https://doi.org/10.1103/t5dh-rx6w} {\bibfield  {journal} {\bibinfo  {journal} {Phys. Rev. Lett.}\ }\textbf {\bibinfo {volume} {135}},\ \bibinfo {pages} {023401} (\bibinfo {year} {2025})}\BibitemShut {NoStop}%
\bibitem [{\citenamefont {Chevy}\ \emph {et~al.}(2002)\citenamefont {Chevy}, \citenamefont {Bretin}, \citenamefont {Rosenbusch}, \citenamefont {Madison},\ and\ \citenamefont {Dalibard}}]{chevy.2002.transverse}%
  \BibitemOpen
  \bibfield  {author} {\bibinfo {author} {\bibfnamefont {F.}~\bibnamefont {Chevy}}, \bibinfo {author} {\bibfnamefont {V.}~\bibnamefont {Bretin}}, \bibinfo {author} {\bibfnamefont {P.}~\bibnamefont {Rosenbusch}}, \bibinfo {author} {\bibfnamefont {K.~W.}\ \bibnamefont {Madison}},\ and\ \bibinfo {author} {\bibfnamefont {J.}~\bibnamefont {Dalibard}},\ }\bibfield  {title} {\bibinfo {title} {Transverse breathing mode of an elongated {Bose}-{Einstein} condensate},\ }\href {https://doi.org/10.1103/PhysRevLett.88.250402} {\bibfield  {journal} {\bibinfo  {journal} {Phys. Rev. Lett.}\ }\textbf {\bibinfo {volume} {88}},\ \bibinfo {pages} {250402} (\bibinfo {year} {2002})}\BibitemShut {NoStop}%
\bibitem [{\citenamefont {Jackson}\ and\ \citenamefont {Zaremba}(2002)}]{jackson.accidental.2002}%
  \BibitemOpen
  \bibfield  {author} {\bibinfo {author} {\bibfnamefont {B.}~\bibnamefont {Jackson}}\ and\ \bibinfo {author} {\bibfnamefont {E.}~\bibnamefont {Zaremba}},\ }\bibfield  {title} {\bibinfo {title} {Accidental suppression of {Landau} damping of the transverse breathing mode in elongated {Bose}-{Einstein} condensates},\ }\href {https://doi.org/10.1103/PhysRevLett.89.150402} {\bibfield  {journal} {\bibinfo  {journal} {Phys. Rev. Lett.}\ }\textbf {\bibinfo {volume} {89}},\ \bibinfo {pages} {150402} (\bibinfo {year} {2002})}\BibitemShut {NoStop}%
\bibitem [{\citenamefont {Pitaevskii}\ and\ \citenamefont {Stringari}(1998)}]{pitaevskii.1998.elementary}%
  \BibitemOpen
  \bibfield  {author} {\bibinfo {author} {\bibfnamefont {L.}~\bibnamefont {Pitaevskii}}\ and\ \bibinfo {author} {\bibfnamefont {S.}~\bibnamefont {Stringari}},\ }\bibfield  {title} {\bibinfo {title} {Elementary excitations in trapped {Bose}-{Einstein} condensed gases beyond the mean-field approximation},\ }\href {https://doi.org/10.1103/PhysRevLett.81.4541} {\bibfield  {journal} {\bibinfo  {journal} {Phys. Rev. Lett.}\ }\textbf {\bibinfo {volume} {81}},\ \bibinfo {pages} {4541} (\bibinfo {year} {1998})}\BibitemShut {NoStop}%
\bibitem [{\citenamefont {Lopes}\ \emph {et~al.}(2015)\citenamefont {Lopes}, \citenamefont {Imanaliev}, \citenamefont {Aspect}, \citenamefont {Cheneau}, \citenamefont {Boiron},\ and\ \citenamefont {Westbrook}}]{lopes.2015.atomic}%
  \BibitemOpen
  \bibfield  {author} {\bibinfo {author} {\bibfnamefont {R.}~\bibnamefont {Lopes}}, \bibinfo {author} {\bibfnamefont {A.}~\bibnamefont {Imanaliev}}, \bibinfo {author} {\bibfnamefont {A.}~\bibnamefont {Aspect}}, \bibinfo {author} {\bibfnamefont {M.}~\bibnamefont {Cheneau}}, \bibinfo {author} {\bibfnamefont {D.}~\bibnamefont {Boiron}},\ and\ \bibinfo {author} {\bibfnamefont {C.~I.}\ \bibnamefont {Westbrook}},\ }\bibfield  {title} {\bibinfo {title} {Atomic {Hong}–{Ou}–{Mandel} experiment},\ }\href {https://doi.org/10.1038/nature14331} {\bibfield  {journal} {\bibinfo  {journal} {Nature}\ }\textbf {\bibinfo {volume} {520}},\ \bibinfo {pages} {66} (\bibinfo {year} {2015})}\BibitemShut {NoStop}%
\bibitem [{\citenamefont {Leprince}\ \emph {et~al.}(2025)\citenamefont {Leprince}, \citenamefont {Gondret}, \citenamefont {Lamirault}, \citenamefont {Dias}, \citenamefont {Marolleau}, \citenamefont {Boiron},\ and\ \citenamefont {Westbrook}}]{leprince.2024.coherent}%
  \BibitemOpen
  \bibfield  {author} {\bibinfo {author} {\bibfnamefont {C.}~\bibnamefont {Leprince}}, \bibinfo {author} {\bibfnamefont {V.}~\bibnamefont {Gondret}}, \bibinfo {author} {\bibfnamefont {C.}~\bibnamefont {Lamirault}}, \bibinfo {author} {\bibfnamefont {R.}~\bibnamefont {Dias}}, \bibinfo {author} {\bibfnamefont {Q.}~\bibnamefont {Marolleau}}, \bibinfo {author} {\bibfnamefont {D.}~\bibnamefont {Boiron}},\ and\ \bibinfo {author} {\bibfnamefont {C.~I.}\ \bibnamefont {Westbrook}},\ }\bibfield  {title} {\bibinfo {title} {Coherent coupling of momentum states: Selectivity and phase control},\ }\href {https://doi.org/10.1103/PhysRevA.111.063304} {\bibfield  {journal} {\bibinfo  {journal} {Phys. Rev. A}\ }\textbf {\bibinfo {volume} {111}},\ \bibinfo {pages} {063304} (\bibinfo {year} {2025})}\BibitemShut {NoStop}%
\bibitem [{\citenamefont {Micheli}\ and\ \citenamefont {Robertson}(2022)}]{micheli.2022.phonon}%
  \BibitemOpen
  \bibfield  {author} {\bibinfo {author} {\bibfnamefont {A.}~\bibnamefont {Micheli}}\ and\ \bibinfo {author} {\bibfnamefont {S.}~\bibnamefont {Robertson}},\ }\bibfield  {title} {\bibinfo {title} {Phonon decay in one-dimensional atomic {Bose} quasicondensates via {Beliaev}-{Landau} damping},\ }\href {https://doi.org/10.1103/PhysRevB.106.214528} {\bibfield  {journal} {\bibinfo  {journal} {Phys. Rev. B}\ }\textbf {\bibinfo {volume} {106}},\ \bibinfo {pages} {214528} (\bibinfo {year} {2022})}\BibitemShut {NoStop}%
\bibitem [{\citenamefont {Butera}\ \emph {et~al.}(2021)\citenamefont {Butera}, \citenamefont {Cl\'ement},\ and\ \citenamefont {Carusotto}}]{butera.2021.position}%
  \BibitemOpen
  \bibfield  {author} {\bibinfo {author} {\bibfnamefont {S.}~\bibnamefont {Butera}}, \bibinfo {author} {\bibfnamefont {D.}~\bibnamefont {Cl\'ement}},\ and\ \bibinfo {author} {\bibfnamefont {I.}~\bibnamefont {Carusotto}},\ }\bibfield  {title} {\bibinfo {title} {Position- and momentum-space two-body correlations in a weakly interacting trapped condensate},\ }\href {https://doi.org/10.1103/PhysRevA.103.013302} {\bibfield  {journal} {\bibinfo  {journal} {Phys. Rev. A}\ }\textbf {\bibinfo {volume} {103}},\ \bibinfo {pages} {013302} (\bibinfo {year} {2021})}\BibitemShut {NoStop}%
\bibitem [{\citenamefont {Gerbier}(2004)}]{gerbier.quasi1d.2004}%
  \BibitemOpen
  \bibfield  {author} {\bibinfo {author} {\bibfnamefont {F.}~\bibnamefont {Gerbier}},\ }\bibfield  {title} {\bibinfo {title} {Quasi-{1D} {Bose}-{Einstein} condensates in the dimensional crossover regime},\ }\href {https://doi.org/10.1209/epl/i2004-10035-7} {\bibfield  {journal} {\bibinfo  {journal} {Europhysics Letters (EPL)}\ }\textbf {\bibinfo {volume} {66}},\ \bibinfo {pages} {771} (\bibinfo {year} {2004})}\BibitemShut {NoStop}%
\bibitem [{\citenamefont {Salasnich}\ \emph {et~al.}(2002)\citenamefont {Salasnich}, \citenamefont {Parola},\ and\ \citenamefont {Reatto}}]{salasnich.2002.effective}%
  \BibitemOpen
  \bibfield  {author} {\bibinfo {author} {\bibfnamefont {L.}~\bibnamefont {Salasnich}}, \bibinfo {author} {\bibfnamefont {A.}~\bibnamefont {Parola}},\ and\ \bibinfo {author} {\bibfnamefont {L.}~\bibnamefont {Reatto}},\ }\bibfield  {title} {\bibinfo {title} {Effective wave equations for the dynamics of cigar-shaped and disk-shaped {Bose} condensates},\ }\href {https://doi.org/10.1103/PhysRevA.65.043614} {\bibfield  {journal} {\bibinfo  {journal} {Phys. Rev. A}\ }\textbf {\bibinfo {volume} {65}},\ \bibinfo {pages} {043614} (\bibinfo {year} {2002})}\BibitemShut {NoStop}%
\bibitem [{\citenamefont {Tozzo}\ and\ \citenamefont {Dalfovo}(2004)}]{tozzo.2004.phonon}%
  \BibitemOpen
  \bibfield  {author} {\bibinfo {author} {\bibfnamefont {C.}~\bibnamefont {Tozzo}}\ and\ \bibinfo {author} {\bibfnamefont {F.}~\bibnamefont {Dalfovo}},\ }\bibfield  {title} {\bibinfo {title} {Phonon evaporation in freely expanding {Bose}-{Einstein} condensates},\ }\href {https://doi.org/10.1103/PhysRevA.69.053606} {\bibfield  {journal} {\bibinfo  {journal} {Phys. Rev. A}\ }\textbf {\bibinfo {volume} {69}},\ \bibinfo {pages} {053606} (\bibinfo {year} {2004})}\BibitemShut {NoStop}%
\bibitem [{Note1()}]{Note1}%
  \BibitemOpen
  \bibinfo {note} {We believe this anisotropy to be small because both radii breathe at the same frequency.}\BibitemShut {Stop}%
\bibitem [{\citenamefont {Micheli}\ and\ \citenamefont {Robertson}(2024)}]{micheli.2024.dissipative}%
  \BibitemOpen
  \bibfield  {author} {\bibinfo {author} {\bibfnamefont {A.}~\bibnamefont {Micheli}}\ and\ \bibinfo {author} {\bibfnamefont {S.}~\bibnamefont {Robertson}},\ }\bibfield  {title} {\bibinfo {title} {Dissipative parametric resonance in a modulated {1D} {Bose} gas},\ }\bibfield  {journal} {\bibinfo  {journal} {Comptes Rendus. Physique}\ }\textbf {\bibinfo {volume} {25}},\ \href {https://doi.org/10.5802/crphys.250} {10.5802/crphys.250} (\bibinfo {year} {2024})\BibitemShut {NoStop}%
\bibitem [{Note2()}]{Note2}%
  \BibitemOpen
  \bibinfo {note} {The Bogoliubov coefficients are given by $u_k, \protect \, v_{k} = \protect \frac {1}{2}\left ( \protect \sqrt {\hbar k^2/(2m\omega _k)} \pm \protect \sqrt {2\omega _km/(\hbar k^2)} \right )$. This transformation depends on the quasiparticle frequency $\omega _k$, which itself depends on the interaction term $g_1n_1$, see Eq.~(\ref {eq:BdG.bk}).}\BibitemShut {Stop}%
\bibitem [{\citenamefont {Tung}\ \emph {et~al.}(2010)\citenamefont {Tung}, \citenamefont {Lamporesi}, \citenamefont {Lobser}, \citenamefont {Xia},\ and\ \citenamefont {Cornell}}]{tung.2010.observation}%
  \BibitemOpen
  \bibfield  {author} {\bibinfo {author} {\bibfnamefont {S.}~\bibnamefont {Tung}}, \bibinfo {author} {\bibfnamefont {G.}~\bibnamefont {Lamporesi}}, \bibinfo {author} {\bibfnamefont {D.}~\bibnamefont {Lobser}}, \bibinfo {author} {\bibfnamefont {L.}~\bibnamefont {Xia}},\ and\ \bibinfo {author} {\bibfnamefont {E.~A.}\ \bibnamefont {Cornell}},\ }\bibfield  {title} {\bibinfo {title} {Observation of the presuperfluid regime in a two-dimensional {Bose} gas},\ }\href {https://doi.org/10.1103/PhysRevLett.105.230408} {\bibfield  {journal} {\bibinfo  {journal} {Phys. Rev. Lett.}\ }\textbf {\bibinfo {volume} {105}},\ \bibinfo {pages} {230408} (\bibinfo {year} {2010})}\BibitemShut {NoStop}%
\bibitem [{\citenamefont {Murthy}\ \emph {et~al.}(2014)\citenamefont {Murthy}, \citenamefont {Kedar}, \citenamefont {Lompe}, \citenamefont {Neidig}, \citenamefont {Ries}, \citenamefont {Wenz}, \citenamefont {Z\"urn},\ and\ \citenamefont {Jochim}}]{murthy.2014.matter}%
  \BibitemOpen
  \bibfield  {author} {\bibinfo {author} {\bibfnamefont {P.~A.}\ \bibnamefont {Murthy}}, \bibinfo {author} {\bibfnamefont {D.}~\bibnamefont {Kedar}}, \bibinfo {author} {\bibfnamefont {T.}~\bibnamefont {Lompe}}, \bibinfo {author} {\bibfnamefont {M.}~\bibnamefont {Neidig}}, \bibinfo {author} {\bibfnamefont {M.~G.}\ \bibnamefont {Ries}}, \bibinfo {author} {\bibfnamefont {A.~N.}\ \bibnamefont {Wenz}}, \bibinfo {author} {\bibfnamefont {G.}~\bibnamefont {Z\"urn}},\ and\ \bibinfo {author} {\bibfnamefont {S.}~\bibnamefont {Jochim}},\ }\bibfield  {title} {\bibinfo {title} {Matter-wave {Fourier} optics with a strongly interacting two-dimensional {Fermi} gas},\ }\href {https://doi.org/10.1103/PhysRevA.90.043611} {\bibfield  {journal} {\bibinfo  {journal} {Phys. Rev. A}\ }\textbf {\bibinfo {volume} {90}},\ \bibinfo {pages} {043611} (\bibinfo {year} {2014})}\BibitemShut {NoStop}%
\bibitem [{\citenamefont {\'Alvarez-Dom\'{\i}nguez}\ and\ \citenamefont {Parra-L\'opez}(2025)}]{alvaro.2025.relevance}%
  \BibitemOpen
  \bibfield  {author} {\bibinfo {author} {\bibfnamefont {A.}~\bibnamefont {\'Alvarez-Dom\'{\i}nguez}}\ and\ \bibinfo {author} {\bibfnamefont {A.}~\bibnamefont {Parra-L\'opez}},\ }\bibfield  {title} {\bibinfo {title} {Relevance of on and off transitions in quantum pair production experiments},\ }\href {https://doi.org/10.1103/rjgn-fp9j} {\bibfield  {journal} {\bibinfo  {journal} {Phys. Rev. D}\ }\textbf {\bibinfo {volume} {112}},\ \bibinfo {pages} {085028} (\bibinfo {year} {2025})}\BibitemShut {NoStop}%
\bibitem [{\citenamefont {Tenart}\ \emph {et~al.}(2021)\citenamefont {Tenart}, \citenamefont {Hercé}, \citenamefont {Bureik}, \citenamefont {Dareau},\ and\ \citenamefont {Clément}}]{tenart.2021.observation}%
  \BibitemOpen
  \bibfield  {author} {\bibinfo {author} {\bibfnamefont {A.}~\bibnamefont {Tenart}}, \bibinfo {author} {\bibfnamefont {G.}~\bibnamefont {Hercé}}, \bibinfo {author} {\bibfnamefont {J.-P.}\ \bibnamefont {Bureik}}, \bibinfo {author} {\bibfnamefont {A.}~\bibnamefont {Dareau}},\ and\ \bibinfo {author} {\bibfnamefont {D.}~\bibnamefont {Clément}},\ }\bibfield  {title} {\bibinfo {title} {Observation of pairs of atoms at opposite momenta in an equilibrium interacting {Bose} gas},\ }\href {https://doi.org/10.1038/s41567-021-01381-2} {\bibfield  {journal} {\bibinfo  {journal} {Nature Physics}\ }\textbf {\bibinfo {volume} {17}},\ \bibinfo {pages} {1364} (\bibinfo {year} {2021})}\BibitemShut {NoStop}%
\bibitem [{\citenamefont {Navon}\ \emph {et~al.}(2021)\citenamefont {Navon}, \citenamefont {Smith},\ and\ \citenamefont {Hadzibabic}}]{navon_quantum_2021}%
  \BibitemOpen
  \bibfield  {author} {\bibinfo {author} {\bibfnamefont {N.}~\bibnamefont {Navon}}, \bibinfo {author} {\bibfnamefont {R.~P.}\ \bibnamefont {Smith}},\ and\ \bibinfo {author} {\bibfnamefont {Z.}~\bibnamefont {Hadzibabic}},\ }\bibfield  {title} {\bibinfo {title} {Quantum {Gases} in {Optical} {Boxes}},\ }\href {https://doi.org/10.1038/s41567-021-01403-z} {\bibfield  {journal} {\bibinfo  {journal} {Nature Physics}\ }\textbf {\bibinfo {volume} {17}},\ \bibinfo {pages} {1334} (\bibinfo {year} {2021})}\BibitemShut {NoStop}%
\bibitem [{Note3()}]{Note3}%
  \BibitemOpen
  \bibinfo {note} {By varying the excitation frequency of the laser power, we could in principle excite other modes~\cite {jaskula.2012.acoustic}. However, the procedure did not yield such a clear parametric growth as in Fig.~\ref {fig2}.}\BibitemShut {Stop}%
\bibitem [{\citenamefont {Gondret}\ \emph {et~al.}(2025{\natexlab{b}})\citenamefont {Gondret}, \citenamefont {Lamirault}, \citenamefont {Dias}, \citenamefont {Leprince}, \citenamefont {Westbrook}, \citenamefont {Cl\'ement},\ and\ \citenamefont {Boiron}}]{gondret.2025.quantifying}%
  \BibitemOpen
  \bibfield  {author} {\bibinfo {author} {\bibfnamefont {V.}~\bibnamefont {Gondret}}, \bibinfo {author} {\bibfnamefont {C.}~\bibnamefont {Lamirault}}, \bibinfo {author} {\bibfnamefont {R.}~\bibnamefont {Dias}}, \bibinfo {author} {\bibfnamefont {C.}~\bibnamefont {Leprince}}, \bibinfo {author} {\bibfnamefont {C.~I.}\ \bibnamefont {Westbrook}}, \bibinfo {author} {\bibfnamefont {D.}~\bibnamefont {Cl\'ement}},\ and\ \bibinfo {author} {\bibfnamefont {D.}~\bibnamefont {Boiron}},\ }\bibfield  {title} {\bibinfo {title} {Quantifying two-mode entanglement of bosonic {Gaussian} states from their full counting statistics},\ }\href {https://doi.org/10.1103/1y1p-zqhh} {\bibfield  {journal} {\bibinfo  {journal} {Phys. Rev. Lett.}\ }\textbf {\bibinfo {volume} {135}},\ \bibinfo {pages} {100201} (\bibinfo {year} {2025}{\natexlab{b}})}\BibitemShut {NoStop}%
\bibitem [{\citenamefont {Menotti}\ and\ \citenamefont {Stringari}(2002)}]{menotti.2002.collective}%
  \BibitemOpen
  \bibfield  {author} {\bibinfo {author} {\bibfnamefont {C.}~\bibnamefont {Menotti}}\ and\ \bibinfo {author} {\bibfnamefont {S.}~\bibnamefont {Stringari}},\ }\bibfield  {title} {\bibinfo {title} {Collective oscillations of a one-dimensional trapped {Bose}-{Einstein} gas},\ }\href {https://doi.org/10.1103/PhysRevA.66.043610} {\bibfield  {journal} {\bibinfo  {journal} {Phys. Rev. A}\ }\textbf {\bibinfo {volume} {66}},\ \bibinfo {pages} {043610} (\bibinfo {year} {2002})}\BibitemShut {NoStop}%
\bibitem [{Note4()}]{Note4}%
  \BibitemOpen
  \bibinfo {note} {Even without assuming a Gaussian profile for the transverse density, a similar result can be obtained~\cite {robertson.2017.controlling}.}\BibitemShut {Stop}%
\bibitem [{\citenamefont {Bogoliubov}(1947)}]{bogoliubov.1947.theory}%
  \BibitemOpen
  \bibfield  {author} {\bibinfo {author} {\bibfnamefont {N.}~\bibnamefont {Bogoliubov}},\ }\bibfield  {title} {\bibinfo {title} {On the theory of superfluidity},\ }\href {https://ufn.ru/pdf/jphysussr/1947/11_1/3jphysussr19471101.pdf} {\bibfield  {journal} {\bibinfo  {journal} {Journal of Physics}\ }\textbf {\bibinfo {volume} {XI}},\ \bibinfo {pages} {23} (\bibinfo {year} {1947})}\BibitemShut {NoStop}%
\bibitem [{\citenamefont {Pitaevskii}\ and\ \citenamefont {Stringari}(2016)}]{pitaevskii.2016.bose_einstein}%
  \BibitemOpen
  \bibfield  {author} {\bibinfo {author} {\bibfnamefont {L.}~\bibnamefont {Pitaevskii}}\ and\ \bibinfo {author} {\bibfnamefont {S.}~\bibnamefont {Stringari}},\ }\href {https://doi.org/10.1093/acprof:oso/9780198758884.001.0001} {\emph {\bibinfo {title} {Bose-Einstein Condensation and Superfluidity}}}\ (\bibinfo  {publisher} {Oxford University Press},\ \bibinfo {year} {2016})\BibitemShut {NoStop}%
\bibitem [{Note5()}]{Note5}%
  \BibitemOpen
  \bibinfo {note} {The difficulties with Bogoliubov treatment for a 1D homogenous gas~\cite {castinmora} are irrelevant here since we focus on a pair of modes only.}\BibitemShut {Stop}%
\bibitem [{\citenamefont {Mora}\ and\ \citenamefont {Castin}(2003)}]{castinmora}%
  \BibitemOpen
  \bibfield  {author} {\bibinfo {author} {\bibfnamefont {C.}~\bibnamefont {Mora}}\ and\ \bibinfo {author} {\bibfnamefont {Y.}~\bibnamefont {Castin}},\ }\bibfield  {title} {\bibinfo {title} {Extension of {Bogoliubov} theory to quasicondensates},\ }\href {https://doi.org/10.1103/PhysRevA.67.053615} {\bibfield  {journal} {\bibinfo  {journal} {Phys. Rev. A}\ }\textbf {\bibinfo {volume} {67}},\ \bibinfo {pages} {053615} (\bibinfo {year} {2003})}\BibitemShut {NoStop}%
\end{thebibliography}%

\end{document}